\documentclass[twocolumn,floatfix,showpacs,superscriptaddress, aps,prb]{revtex4-1}
\usepackage{amsmath,amsfonts,amssymb}
\usepackage{graphicx}
\usepackage{bm}
\usepackage[utf8]{inputenc}

\begin{document}
\title{Probing the core and shell contributions to exchange bias in Co/Co$_3$O$_4$ nanoparticles of controlled size}

\author{D. De}
\affiliation{Department of Solid State Physics, Indian Association for the Cultivation of Science, Jadavpur, Kolkata 700032, India}
\affiliation{Department of Physics, The Neotia University, D. H. Road, 24 PGS (S), W.B. 743363, India}
\author{\`{O}scar  Iglesias}
\email{oscar@ffn.ub.es}
\homepage{http://www.ffn.ub.es/oscar}
\affiliation{Departament de Física de la Mat\'{e}ria Condensada and Institut de Nanociència i Nanotecnologia (IN$^ 2$UB), Facultat de Física, Universitat de Barcelona, Av. Diagonal 647, 08028 Barcelona, Spain}
\author{S. Majumdar}
\author{Saurav Giri}
\email{sspsg2@iacs.res.in}
\affiliation{Department of Solid State Physics, Indian Association for the Cultivation of Science, Jadavpur, Kolkata 700032, India}

\begin{abstract}
Coupling at the interface of core/shell magnetic nanoparticles is known to be responsible for the exchange bias (EB)  and the relative sizes of core and shell components are supposed to influence  the associated phenomenology.
In this work, we have prepared core/shell structured nanoparticles with the total averaged diameter around $\sim$ 27 nm and a wide range of shell thicknesses through the controlled oxidation of Co nanoparticles well dispersed in an amorphous silica host. Structural characterizations give compelling evidence of the formation of Co$_3$O$_4$  crystallite phase at the shells surrounding the Co core. Field cooled hysteresis loops display nonmonotonous dependence of the exchange bias $H_E$  and coercive $H_C$ fields, that become maximum for a sample with an intermediate shell thickness, at which lattice strain is also maximum for both the phases. Results of our atomistic Monte Carlo simulations of the particles with the same size and compositions as in experiments are in agreement with the experimental observations and have allowed us to identify a change in the contribution of the interfacial surface spins to the magnetization reversal giving rise to the maximum in $H_E$ and $H_C$.

\end{abstract}
\date{\today}

\maketitle

\section{Introduction}

Nanocrystalline materials are of potential interest nowadays, because their remarkable properties, that differ from their parent bulk counterparts,  have found a wide range of technological applications. Among them, particles made of magnetic compounds display a variety of magnetic behaviors, that differ substantially from their parent massive materials. \cite{Dormann_ACP97,Kodama_jmmm1999,Fiorani_book2005} Their distinct properties are mainly connected to the finite size effects related to the reduced number of magnetic ions in the enclosed volume. \cite{Iglesias_prb2001} Additionally, the surface and interface effects related to the symmetry breaking at physical boundaries of the materials cause spin disorder and frustration along with the interparticle interactions. \cite{Sabsabi_prb2013}  An interesting class of nanoparticles (NP) is found when ferromagnetic (FM) and antiferromagnetic (AF) materials are combined together in a core/shell structure. The coupling at the interface between these two magnetic phases gives rise to the EB phenomenon, \cite{Nogues_JMMM1999,Nogues_physrep2005,Iglesias_jnn2008,Giri_jpcm2011,Manna_PhysRep2014} which is of fundamental interest and has found multiple technological applications depending on the specific composition and the characteristic size of the respective phases.  \cite{Gierlings_PRB2002,Skumryev_Nature2003,Velthuis_jap2000,Radu_prb2003,Das_jac2009,Wu_jpcm2007}
The pinning mechanism, that results from EB, has been commercially explored for the magnetic field sensors and in modern magnetic read heads. \cite{Chappert_NatMater2007,Lopez-Ortega_PhysRep2015} Nevertheless, a clear cut connection between the observed EB phenomenology and parameters in core/shell NP, such as the size and thickness of the NP or the microscopic interfacial structure, is not well established yet.

Although pure metal particles would have desirable high values of  saturation of magnetization, \cite{OHandley_2000}  they have strong natural tendency to form parent oxide phases. \cite{Haneda_Nature1979,Giri_ASC2001} This process can be controlled under proper synthesis conditions to prevent further oxidation, leading to  the formation of core/shell structures with the oxide phase (often an AF or ferrimagnetic material \cite{Goodenough_book1963}) usually formed at the outer part of the structures. 
In the case of Co NP, most of the published studies\cite{Meiklejohn_prb56} report the formation of the CoO phase on the shell, although in some cases the presence of  the Co$_3$O$_4$  has also been evidenced  by structural \cite{Fontaina_NL2004,Li_SciRep2015} or magnetic characterization. \cite{Simeonidis_prb2011} 
The possibility of observing EB in Co based nanostructures in contact with Co$_3$O$_4$ has been rather less investigated and the published studies focus on layered structures. \cite{Wang_ijmpb2005,Wang_ssc2005,You_jap2003,Wang_jac2008,Ahmad_jmmm2015}  Synthesis of single phase CoO or Co$_3$O$_4$ NP have been achieved 
by several authors, that have reported AF magnetic behavior with ordering temperatures reduced compared to the bulk values \cite{Bisht_scc2011,Ichiyanagi_poly2005,Dutta_jpcm2008} and remanence values much higher than those for the bulk due to finite-size effects. \cite{Simeonidis_prb2011,Silva_prb2010}

Here, we will explore the EB effect in Co/Co$_3$O$_4$ batches of nanoparticles with well defined average size, giving evidences that Co$_3$O$_4$ is the only phase present in our samples. The sizes of the crystallites forming core and shell are tuned by the controlled oxidation at different temperatures. Hence, we are able to show that, the variation of shell thickness, while keeping the fixed overall particle size, leads to the significant changes in the EB effect, that are stronger for the intermediate shell thicknesses.  A detailed structural study of the samples allows to correlate this maximum EB to the maximum interfacial strain due to lattice mismatch and the associated increased anisotropy. By means of atomistic Monte Carlo (MC) simulations, we trace back the origin of  maximum EB effect at the intermediate size and the changes in the magnetization reversal of interfacial surface spins as core/shell size is varied.
\begin{figure}[tbp]
\vskip 0.0 cm
\centering
\includegraphics[width = 0.95\columnwidth]{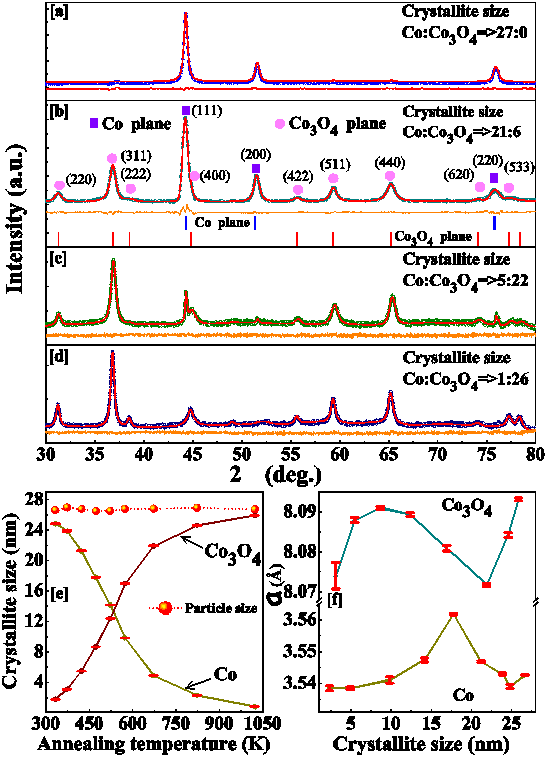}
\caption {(a-d) XRD patterns of Co/Co$_3$O$_4$ nanostructures with different core/shell crystallite sizes. Continuous curves show the fit using Rietveld refinement. Difference plot are shown at the bottom of each pattern. Details of the indexed planes corresponding to Co and Co$_3$O$_4$ are given in (b) for the $21:6$ sample. Variation of the (e) crystallite sizes of the  two phases and overall size of the particles with the annealing temperatures and (f) lattice constant ($a$) of the two phases with the crystallite size, as obtained from the refinement.} 
\label{Fig.1}
\end{figure}

\section{Experimental}
Nanocrystalline Co embedded in the amorphous silica host is prepared with volume fraction $\varphi$=10 \% using sol gel route. Initially, Co metal powder (Aldrich, 99.99 \% pure) is dissolved in nitric acid (4.5 M). Citric acid is then added to the solution and homogenized thoroughly during 6 h to obtain a  transparent reddish solution. Ethanolic tetraethyl orthosilicate (TEOS) is finally added to the solution as a source of the silica matrix in form of droplets and mixed vigorously for 12 h for obtaining a homogeneous solution. The final reddish solution is dried in open air very slowly to form a gel at room temperature, that is  dried at 323 K for 15 days and subsequently decomposed at 873 K for 6 h in a continuous flow of H$_2$/Argon mixture (5\% H$_2$ and 95\% Argon). The as-synthesized Co nanoparticles ($\varphi \sim$ 10 \%) in a silica matrix are processed for controlled oxidation by annealing the sample in the range of 333 - 1023 K for 10 minutes each in order to grow desired Co/Co$_3$O$_4$ phase fractions. Henceforth, nine samples will be addressed as 25:2, 24:3, 21:6, 18:9, 15:12, 10:17, 5:22, 2:25, and 1:26, where numbers are the sizes of Co and Co$_3$O$_4$ phases, respectively in nm. 

Chemical composition is confirmed using powder X-ray diffraction (XRD) studies (Seifert XRD 3000P) considering Cu K$\alpha$ radiation and electron diffraction attached with a Transmission Electron Microscopy (TEM) of JEOL TEM 2010. High resolution TEM images of the particles are used to assess their size, shape and crystalline planes of Co and Co$_3$O$_4$.  X-ray photoemission spectroscopy (XPS) have been performed in an Omicron Nanotechnology spectrometer. Magnetization is recorded in a commercial SQUID magnetometer of Quantum Design (MPMS, XL). In the zero-field cooled (ZFC) protocol the sample is cooled in zero-field and the magnetization is measured in the warming mode with a static magnetic field. In the field-cooled (FC) protocol sample is cooled and measured in field.

\begin{figure}[tbp]
\vskip 0.0 cm
\centering
\includegraphics[width = 0.9\columnwidth]{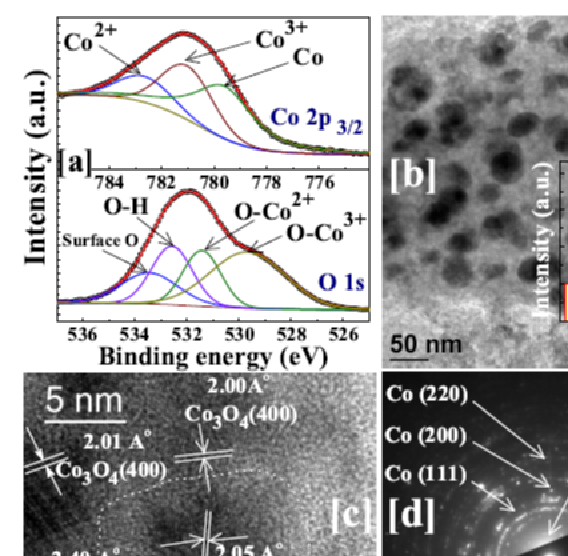}
\caption {(a) XPS spectrograms of Co 2p$_{3/2}$ (top panel) and O 1s (bottom panel) contributions. (b) TEM image to verify particle size. Inset: particle size histogram as fitted by a log-normal distribution function. (c) High resolution TEM image displaying nearly core shell structure composed of Co and Co$_3$O$_4$, respectively. Different planes corresponding to Co and Co$_3$O$_4$ are depicted. (d) Electron diffraction pattern displaying different planes for 15:12 sample.}
\label{Fig.2}
\end{figure}

\section{Structural Characterization}
\label{Sec_Charact}

The XRD patterns of Co and Co/Co$_3$O$_4$ of selected samples with different phase fractions are depicted in Figs. \ref{Fig.1}(a-d). The continuous curves show the corresponding fits obtained by Rietveld refinement with the MAUD software, using face-centered $Fm3m$ and $Fd3m$ space groups for Co and Co$_3$O$_4$, respectively. The details of the refinement are shown in Fig. \ref{Fig.1}(b), where the vertical bars at the bottom depict the diffraction peak positions of Co and Co$_3$O$_4$. The goodness of the fit is demonstrated by the difference plot shown below the diffraction patterns for all the compositions, which confirms absence of other impurity phases such as minor fraction of CoO. With increasing annealing temperature the Co$_3$O$_4$ phase grows at the expense of Co. 

Average crystallite sizes of each component for different annealing temperatures are estimated using the Scherrer formula from broadening of the diffraction peaks, as obtained from the Rietveld refinement. \cite{Cullity_book2004} Individual average crystallite sizes of each component for annealing temperatures in the range $333-1023$ K are depicted in Fig. \ref{Fig.1}(e). 
As shown, the average particle size $\sim$ 27 nm is almost constant  for the different annealing temperatures. 
The lattice constants ($a$) as obtained from the Rietveld refinements are plotted as a function of the individual crystallite sizes in Fig. \ref{Fig.1}(f). The obtained values are consistent with the previously reported results. \cite{Dutta_jpcm2008,Zhuo_jcgd2009}
For most of the samples the lattice constants of Co and Co$_3$O$_4$ deviate from their bulk counterpart values 3.54 and 8.09 \AA, \cite{Nishizawa_bapd1983,Dutta_jpcm2008}  being higher and lower than in bulk, respectively. This reveals significant tensile strains on the Co cores of the particles caused by the formation of the oxide phase at the shell. Interestingly, a maximum value of both $a$ is found for the sample obtained at an annealing temperature of  473 K, 
with  crystallite sizes  $\sim$ 18 and $\sim$ 9 nm for Co and Co$_3$O$_4$, respectively.
%
\begin{figure}[tbp]
\vskip 0.0 cm
\centering
\includegraphics[width = \columnwidth]{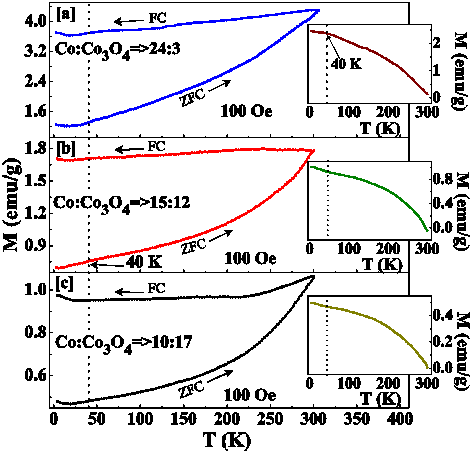}
\caption {Thermal variations of ZFC and FC magnetization for Co:Co$_3$O$_4$ with (a) 24:3, (b) 15:12, and (c) 10:17. The corresponding insets show thermomagnetic irreversibility ($\Delta M$).}
\label{Fig. 3}
\end{figure}
%
 
%
Analysis of the XPS spectra have been performed in order to investigate the details of the chemical composition of the core and shell phases of the nanoparticles.
The results for the Co 2p$_{3/2}$ and O 1s contributions to XPS are depicted in Fig. \ref{Fig.2}(a) for the sample with 15:12 composition. 
The oxidation states of Co atoms are obtained by deconvoluting the spectra for Co 2p$_{3/2}$ and O 1s contributions, as shown at the top and bottom panels of the figure. 
The Co$^{2+}$,  Co$^{3+}$ and Co contributions to the 2p$_{3/2}$ spectrum can be deconvoluted as shown in the corresponding subspectra (lines) that exhibit maxima with increasing binding energies, respectively. 
A similar procedure is done for the O 1s spectrum, that can be deconvoluted into four contributions,  exhibiting maxima with increasing energy corresponding to surface O, O-H,  O-Co$^{2+}$, and O-Co$^{3+}$, respectively.
It is noted that the ratio of the area under the deconvoluted curves of Co$^{3+}$ and Co$^{2+}$ in Co 2p$_{3/2}$ spectrum, as well as that of O-Co$^{3+}$and O-Co$^{2+}$ curves in O 1s spectrum has nearly same value of 2:1, as expected for Co$_3$O$_4$. 
Furthermore, the analysis provides a ratio of Co:(Co$^{2+}$+Co$^{3+}$) $\approx$ 50.3:49.7, which is close to the phase fraction ratio of Co:Co$_3$O$_4 \approx$ 52:48, as obtained from the Rietveld refinement. This two observations clearly corroborate the absence of any detectable contribution from CoO as also indicated by the XRD analysis mentioned above.

Figure \ref{Fig.2}(b) shows the TEM image of the sample 15:12. As depicted in the inset, there is a distribution of particle sizes that can be fitted  using log-normal distribution function with an average size of $\sim$ 24 nm (consistent with that observed from the XRD results) and tails that extend from  5 to 40 nm  (continuous curve).
A high resolution TEM image of a particle of the same sample is shown in Fig. \ref{Fig.2}(c), where we have indicated, the distinct Co and Co$_3$O$_4$  areas within the particle. 
Moreover, we have identified the Co (111) diffraction planes within the darker core region and, outside, the distinctive planes of Co$_3$O$_4$, that could also be observed in the XRD patterns. These representative Co and Co$_3$O$_4$ planes can also be clearly resolved in the electron diffraction pattern shown in Fig. \ref{Fig.2}(d). All in all, the careful structural analysis does not show any convincing signature CoO phase, consistent with the XRD results.

\begin{figure*}[tbp]
\vskip 0.0 cm
\centering
\includegraphics[width = 12 cm]{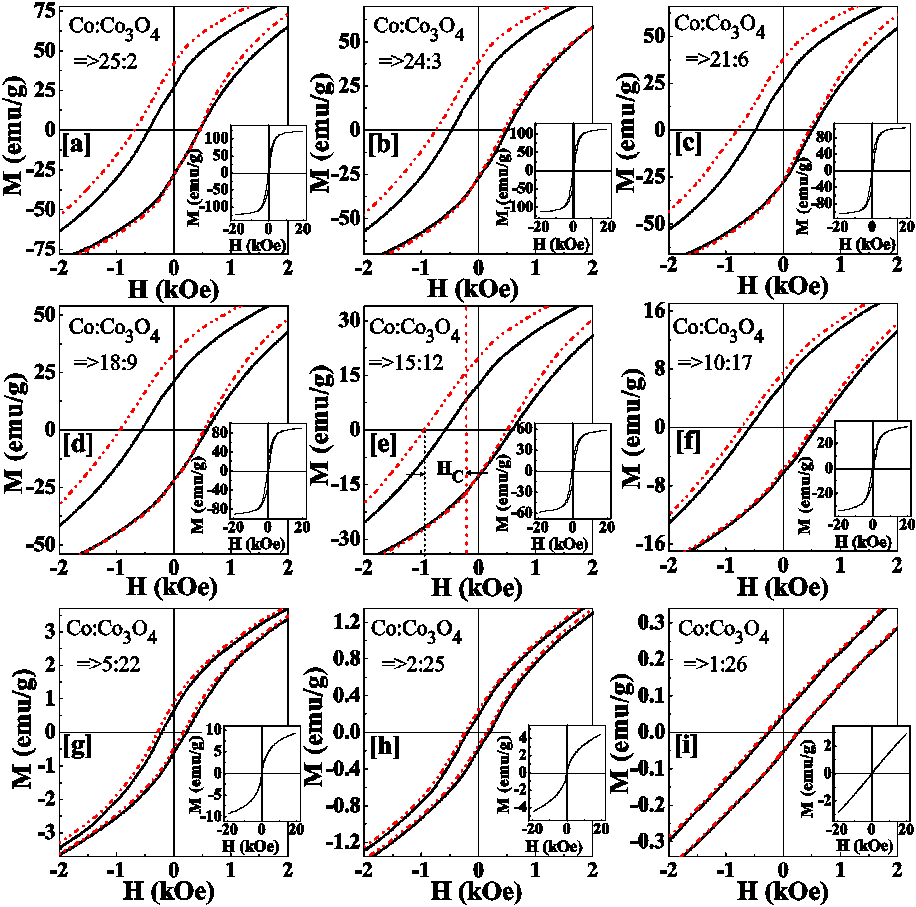}
\caption {Central portion of the field-cooled (dashed curve) and zero field-cooled (continuous curve) magnetic hysteresis loops at 5 K for Co:Co$_3$O$_4$ with crystallite sizes (in nm) (a) 25:2, (b) 24:3, (c) 21:6, (d) 18:9, (e) 15:12, (f) 10:17, (g) 5:22, (h) 2:25, and (i) 1:26. Cooling field for FC protocol is 10 kOe. Complete ZFC hysteresis loops are depicted in the corresponding insets.} 
\label{Fig.4}
\end{figure*}    

\section{Magnetic characterization}
The thermal variation of dc magnetization measured under ZFC and FC protocols with an applied magnetic field of 100 Oe are depicted in Fig. \ref{Fig. 3}(a-c) for samples  24:3, 15:12, and 10:17. The behavior of the three samples is similar, showing  irreversibility up to the highest measured temperature of 300 K, which indicates that the nanoparticles have blocking temperatures above this value due to their relatively large size. Note that both curves decrease monotonously below 300 K and do not display any peak characteristic of the Néel temperature of CoO, which should be in the range of $235-293$ K depending on the particle size. \cite{Chandra_apl2012,Feygenson_prb2010,Simeonidis_prb2011} 
The Co$_3$O$_4$ orders antiferromagnetically around 40 K. \cite{Roth_jpcs1964} Below $\sim$ 40 K, a weak anomaly can be observed in both the ZFC and FC curves marked by an upturn of the magnetization, which could be ascribed to the onset of the antiferromagnetic order \cite{Simeonidis_prb2011} in crystallites forming the shell. This is consistent with the significant decrease of the Néel temperature of Co$_3$O$_4$ in the range of 26$-$35 K depending on the finite size effect. \cite{Dutta_jpcm2008,Dreifus_mre2015}
With increasing Co$_3$O$_4$ fraction, the magnitude of the magnetization as well as the thermomagnetic irreversibility (defined as $\Delta M=M_{FC}-M_{ZFC}$)  decrease, as a result of the increasing contribution of shell spins with reduced magnetization,  as shown in Fig. \ref{Fig. 3} and the corresponding insets. Similar low temperature responses (not to be confused with the ones reported here) have been reported  that are usually ascribed either to residual phases \cite{Simeonidis_prb2011} in Co/CoO nanoparticles or onset of  spin-glass freezing \cite{Chandra_apl2012,Tracy_prb2005} for other core/shell compositions.
 
\begin{figure}[tbp]
\vskip 0.0 cm
\centering
\includegraphics[width =0.95\columnwidth]{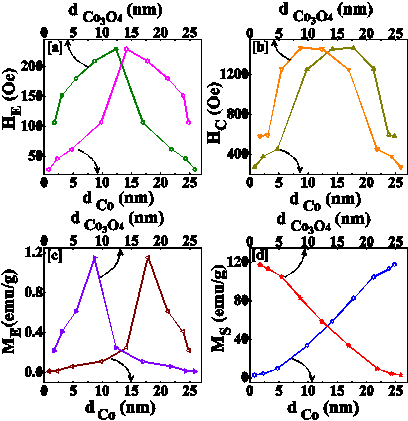}
\caption {Variation of (a) $H_E$, (b) $H_C$, (c) $M_E$, and (d) $M_S$ with crystallite sizes of Co ($d_{Co}$) and Co$_3$O$_4$ ($d_{Co_3O_4}$).} 
\label{Fig.5}
\end{figure}

Hysteresis loops have been measured in between $\pm$ 20 kOe at 5 K for all the nine samples, as shown by the full loops in the insets of Figs. \ref{Fig.4}(a-i). For particles with dominant Co component at the particle core (samples 25:2, 24:3), the hysteresis loop exhibits the typical shape of a FM material, being reversible well below 20 kOe, and the high field magnetization reaches values close to saturation for bulk Co ($\sim 162 $ emu/g, see  Fig. \ref{Fig.5}d).\cite{OHandley_2000}  
With the increase of the oxide component at the shell, the loop shapes become more elongated with higher closure fields \cite{Simeonidis_prb2011,Chandra_apl2012} and a high field linear component with increasing contribution. This high field susceptibility can be ascribed to the contribution of  uncompensated spins in the antiferromagnetic shell  and core/shell interface\cite{Tracy_prb2005,Tracy_prb2006,Rinaldi-Montes_jmcc2016} as it dies off at temperatures higher than T$_N$ of  Co$_3$O$_4$.  For the most oxidized sample (sample 1:26) a linear field dependence extends over the whole field range as typically reported for purely antiferromagnetic nanoparticles or in bulk. \cite{Kovylina_nanot2009,Tracy_prb2005,Dutta_jpcm2008}  

In order to probe the effect of the shell thickness on the EB effect, all the nine samples are cooled in $H_{cool}=$10 kOe from 300 K down to 5 K and the magnetic hysteresis loops are recorded subsequently. The main panels in Figs. \ref{Fig.4}(a-i) show a zoom of the central portion of the hysteresis loops for ZFC (continuous lines) and FC (dashed lines)  protocols. 
Note that hysteresis loops after FC appear clearly shifted to the negative fields with respect to ZFC ones as a consequence of the EB coupling between the FM core and AF shell spins.
In some cases, a vertical displacement is also observed. 
The EB field and vertical shift are defined as $H_{E}=(H_C^++H_C^-)/2$, $M_E=[M(20\, {\rm kOe})+M(-20\, {\rm kOe})]/2$ respectively, where $H_C^+$ and $H_C^-$ are the coercive fields at the decreasing and increasing field loop branches. The dependence of these quantities as well as that of the coercive field $H_C$ and saturation magnetization$M_S$ on the crystallite size of the both phases is  presented in Figs. \ref{Fig.5}(a-d).

The most remarkable feature is the nonmonotonic behavior of $H_E$ and $H_C$ presented in Fig. \ref{Fig.5}, which has been reported previously for Co/CoO nanoparticles. \cite{Gangopadhyay_jap1993,Iglesias_jnn2008,Kovylina_nanot2009,Feygenson_prb2010} It is clear that EB has to tend towards zero when either the oxide shell becomes thinner or the Co core diminishes. However, an argument based on finite size effects would indicate that with increasing the core diameter $H_E$ should increase resulting from the increase of surface interface. On the other hand, a decrease in the shell thickness  would produce the the contrary effect. So, in order to clarify the origin of the nonmonotonic behavior, results of MC simulations will be presented in Sec. \ref{Sec_Sim}. 
However, it should be noted that the maximum $H_E$ and $M_E$ is found for the sample with 18:9 composition, which was the one showing higher lattice mismatch. Therefore, lattice strain at the core/shell interface for the intermediate shell thickness might be playing a role by inducing a higher net magnetic moment at the interface. The result is in agreement with what it has been found at the interface of Au-Fe$_3$O$_4$ composites \cite{Chandra_Nano2014,Feygenson_prb2015} or Co/Co$_3$O$_4$ nanooctahedra, \cite{Li_SciRep2015} suggesting that, lattice mismatch correlates the magnetic anisotropy.  
\begin{figure}[tbp]
\vskip 0.0 cm
\centering
\includegraphics[width = 0.95\columnwidth]{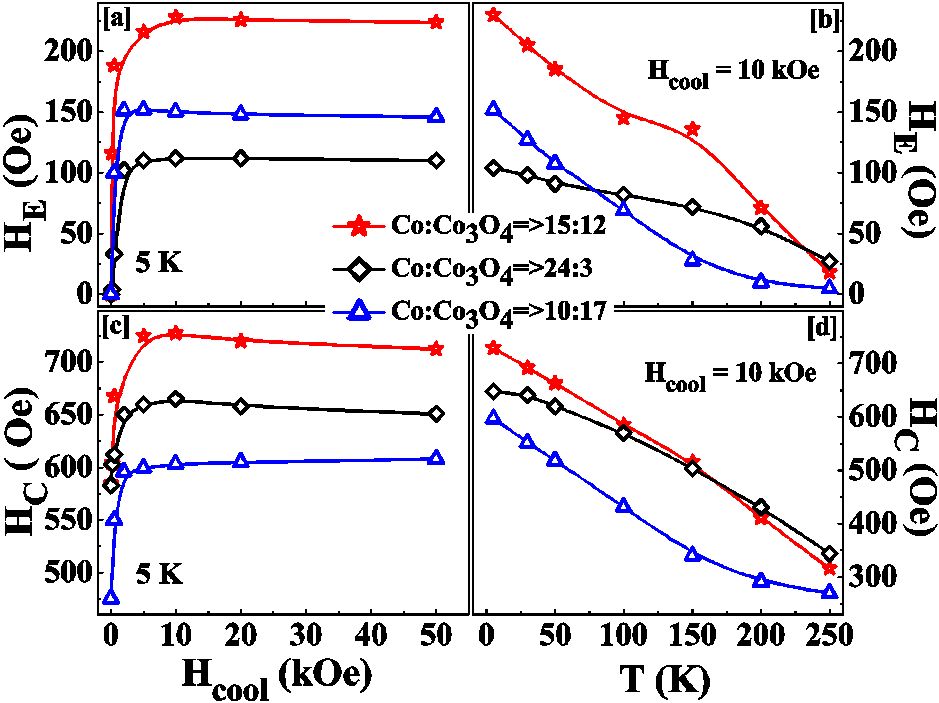}
\caption {Cooling field ($H_{cool}$) dependence of (a) $H_E$ and (c) $H_C$ and temperature variation of (b) $H_E$ and (d) $H_C$ for three selected samples as indicated in the legend.} 
\label{Fig.6}
\end{figure}
This hypothesis is further sustained by the behavior of $H_C$ shown in Fig. \ref{Fig.5}(b), that presents a maximum for sample 18:9, and with the results of the MC simulations presented in Sec. \ref{Sec_Sim}. Finally, we notice that vertical loop shifts indicated in Fig. \ref{Fig.5}(c) are concomitant to the observation of horizontal shifts and indicate the existence of a fraction of spins that remain pinned during the field reversal  (see also the simulation results in Fig.\ref{Fig.7} below). Therefore, the coincidence of the maximum for both the quantities points to a relation between the increased interfacial anisotropy and stress as commented above.

We have also measured the thermal variation of FC hysteresis loops by cooling the sample in a 10 kOe cooling field from 300 K down to the various temperatures below 250 K and the $H_{cool}$ dependence by  cooling the samples from 300 K down to 5 K in different cooling fields up to 50 kOe. The extracted $H_E$ and $H_C$ values and their variation with the mentioned parameters are presented in Fig. \ref{Fig.6} for 24:3, 15:12, and 10:17 samples.
The values of $H_E$ and $H_C$ increase rapidly  with $H_{cool}$ initially and saturate for  $H_{cool} >$ 10 kOe for the three selected samples. Therefore, we can exclude that the loop shifts observed in our samples are due to minor loop effects. No maximum in these quantities has been detected in our samples, in contrast with what is observed in some studies of single phase oxide nanoparticles, \cite{DelBianco_prb2004,Fiorani_jpcm2007,Vasilakaki_prb2009} where it was related to the glassy magnetic nature of surface spins.  
The results displayed in Figs. \ref{Fig.6}(b,d) show that both $H_E$ and $H_C$ decrease with increasing temperature although following different tendencies depending on the sample. Remarkably, although the Néel temperature of Co$_3$O$_4$ is below $40$ K, the loop shifts persist up to $\sim$ 250 K or above, demonstrating the robustness of the exchange coupling between core and shell and the persistence of EB effects up to almost room temperature. 
\begin{figure}[tbp]
\vskip 0.0 cm
\centering
\includegraphics[width = \columnwidth]{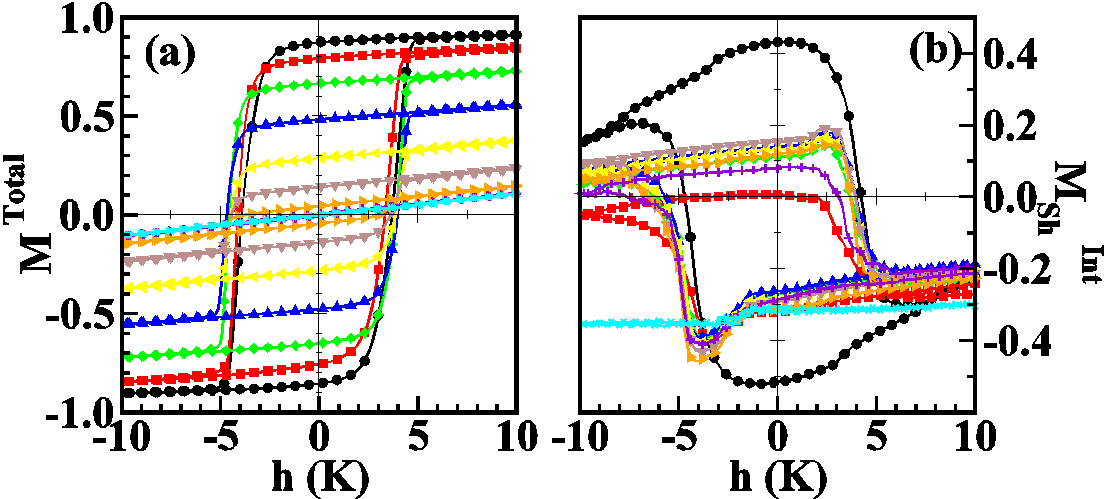}
\caption {Simulated hysteresis loops for individual particles with the same dimensions as experimental samples with increasing shell thickness from outer to innermost loops. Panel (a) shows the normalized magnetization of the whole particle. Panel (b) shows the contribution of the interfacial spins at the shell.}
\label{Fig.7}
\end{figure}

\section{Simulation results}
\label{Sec_Sim}
To sustain the experimentally observed variation of $H_E$ and $H_C$ with particle morphology, we have conducted atomistic MC simulations of a model of core/shell nanoparticle \cite{Cabot_prb2009,Simeonidis_prb2011} based on the following  Hamiltonian
\begin{equation}
{\cal H}=-\sum_{< i,j>} J_{ij}\vec{S_i} \cdot \vec{S_j}
-\sum_{i}  K_i \left(\vec{S_i} \cdot \hat{n_i}\right)^2
-\sum_{i}\vec{S_i} \vec{h}  \ ,
\label{Eq1}
\end{equation}
where $\vec{S}_i$ are Heisenberg classical unit vectors representing the magnetic ions and, in the last term,  the magnitude of the magnetic field $\vec{H}$ is given in reduced units $h=\mu_S H$, with $\mu_S$ the atomic spin moment. Note also that all parameters used in the simulations will be given in temperature units, scaling them by the Boltzmann constant $k_B$ and that the simulation temperature  includes a factor $1/S^2$.
The real values of the exchange and anisotropy  constants for Co and Co$_3$O$_4$ have been used \cite{Balcells_apl09} and they are given  by $J_{C}= 92.75$ K, $J_{Sh}= 0.23 J_C$ and $K_{Sh}= 40.1 J_C, K_{C}= 0.022 J_C$. The interface exchange coupling has been set to $J_{int}=-J_{C}$.  The total radius of the simulated particles (containing $212095$ spins) has been taken as the mean radius of the real samples $R=38 a$ (where $a$  is the lattice constant) and nine shell thicknesses $t_{sh}/a=2.4, 4.4 ,7.7, 12.3, 17.5, 23.8, 30.8, 34.6, 36.5$ have been considered as studied experimentally. In order to mimick the observed presence of crystallites in the shell of the real nanoparticles (see Sec. \ref{Sec_Charact}), we have divided the shell in regions with different random anisotropy directions similar to that was done in the literature. \cite{Cabot_prb2009,Simeonidis_prb2011} This turns out to be crucial to reproduce the experimental phenomenology. 

In Fig. \ref{Fig.7}(a) we present the simulated hysteresis loops as obtained after cooling down to 0.1 K under an applied magnetic field, $h_{FC}=10$ K, for different shell thicknesses. The loops display shifts contrary to the cooling field direction, that vary in a non-monotonous way with $t_{sh}$. With increasing  $t_{sh}$  the fraction of AF spins increases. As a result of it, the remanent magnetization decreases and the loops become more similar to that of an AF material. The area of the loops decrease with an extended region having linear field dependence, in qualitative agreement with the experimentally observed behavior.
\begin{figure}[tbp]
\vskip 0.0 cm
\centering
\includegraphics[width = \columnwidth]{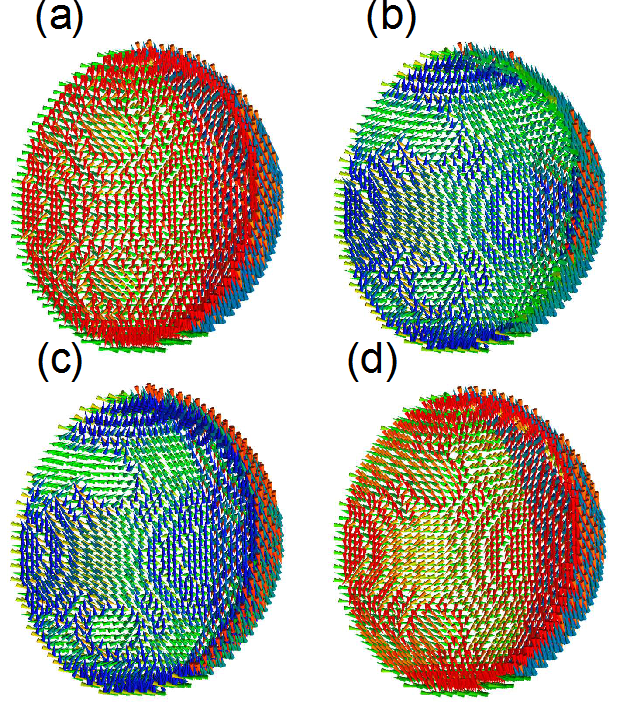}
\caption {Snapshots of a slice of a nanoparticle with $t_{sh}/a=17.5$  (sample with 15:12 composition) showing the interfacial spin magnetic configurations at different points of the hysteresis loop (from left to right): (a) positive remanence point, (b) negative coercive field, (c) negative remanent point, and (d) positive coercive field. Cone colors vary depending their component along the field direction from red (along the field direction) to blue (contrary to the field direction) following the visible light spectrum.}  
\label{Fig.8}
\end{figure}

The variation of the EB field can be more easily traced back to the magnetization reversal behavior of the interfacial spins at the shell, \cite{Iglesias_prb2005} whose contribution to the hysteresis loop is shown in Fig. \ref{Fig.7}(b). As can be clearly seen, the interfacial  hysteresis loops present a clear asymmetry between the decreasing and increasing field branches and, some of them, display characteristic apexes near the coercive field points.\cite{Iglesias_prb2005,Wu_jpcm2007} Although the interface magnetization remains quite constant except near the coercive fields, the interfacial magnetization on the decreasing field branch is not equal (in absolute value) to that on the increasing field. This indicates that a considerable fraction of interfacial spins remains pinned during the field reversal.  
\begin{figure}[tbp]
\vskip 0.0 cm
\centering
\includegraphics[width = \columnwidth]{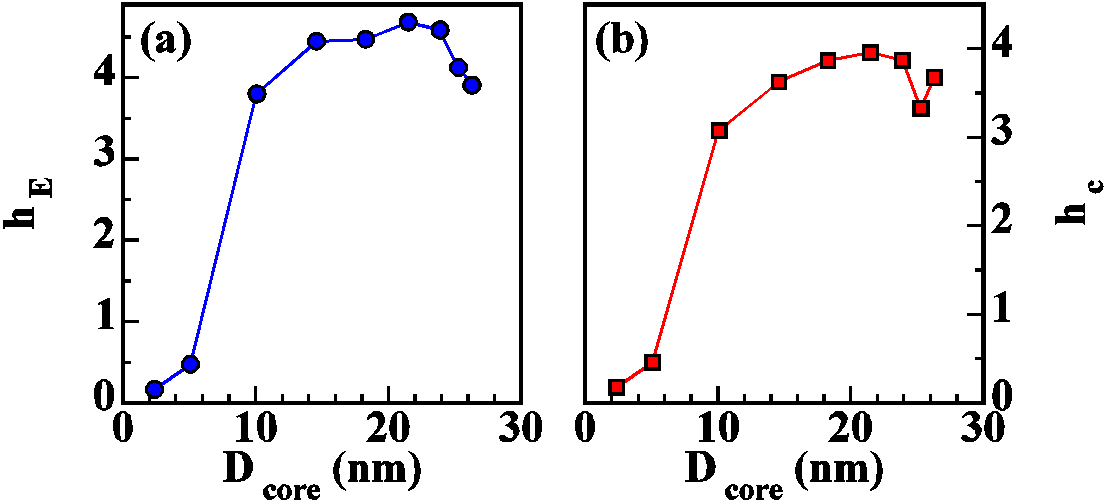}
\includegraphics[width = \columnwidth]{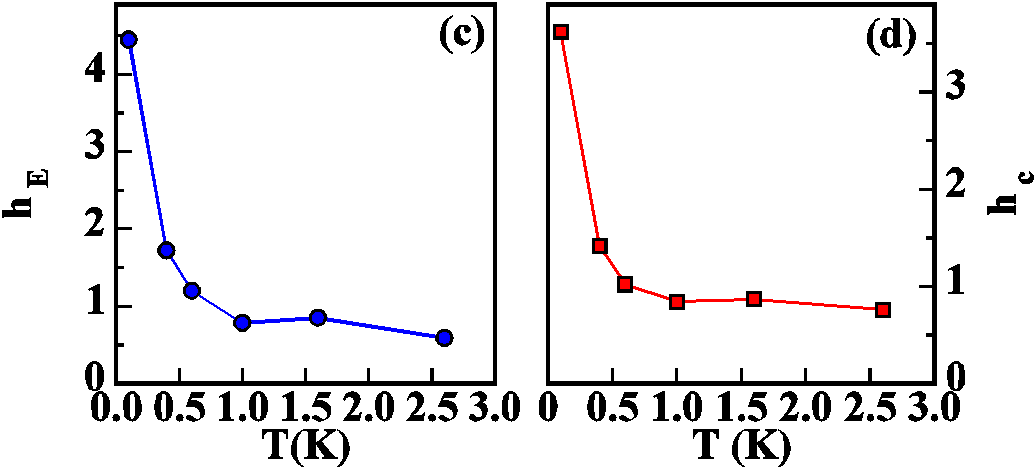}
\caption {Upper panels show the variation of (a) the shift $h_{E}$  and (b) coercive field $h_C$ of the simulated hysteresis loops with the diameter of the particle core. Lower panels show the thermal dependence of  (c) $h_{E}$  and (d) $h_C$  for a particle with shell size $t_{shell}= 17a$.}
\label{Fig.9}
\end{figure}
This can be directly checked by looking at snapshots of the interfacial spins configurations taken at different points of the hysteresis loops, as displayed in Fig. \ref{Fig.8} for a particle with $t_{sh}/a=17.5$.  Comparing the magnetic configurations at the remanent [panels (a) and (c)] and coercive field points [panels (b) and (d)], we can see that the interfacial surface spins  remain mostly oriented along the direction induced by the magnetic field applied during the initial cooling, as indicated by the absence of variation in the colors of the outer shell of spins. On the contrary, interfacial spins in contact with the core, are dragged during the quasiuniform core reversal as can be appreciated by change in color (reddish to blueish and vice versa) and orientation when going from remanent to the coercive field points.

In order to compare with the experimental results of Fig. \ref{Fig.5} and \ref{Fig.6}, we have calculated the coercive field and horizontal loop shifts as $h_C=(h_c^+-h_c^-)/2$, $h_{E}=(h_c^++h_c^-)/2$ from the hysteresis loops of Fig. \ref{Fig.7}. Their dependence on the core diameter is given in Fig. \ref{Fig.9}.  Initially, both the quantities increase with increasing the particle core diameter starting from a fully AF particle corresponding to the increase of the interfacial region surface, as it is also observed experimentally. However, this tendency is broken as the core size increases further and a maximum in $h_{E}$  is observed for a core diameter of $20$ nm (shell thickness of $6$ nm). Below this value, the EB field progressively decreases as the shell thickness is reduced. This non-monotonous trend is in agreement with that observed experimentally. 

\begin{figure}[tbp]
\vskip 0.0 cm
\centering
\includegraphics[width = 0.6\columnwidth]{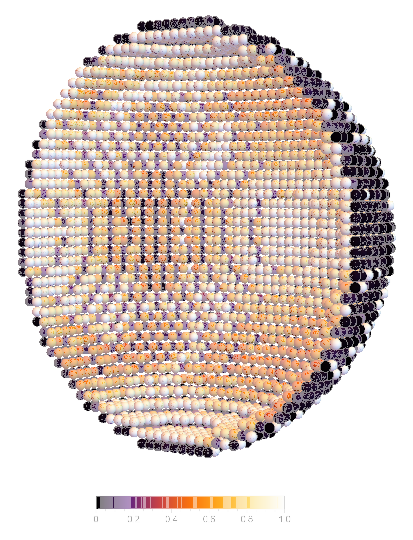}
\caption {Snapshot showing a cut of the interfacial spin positions represented by spheres. Sphere coloring varies with the  normalized magnitude of the difference between local spin orientations at the two coercive field points. Lighter color means more difference.} 
\label{Fig.10}
\end{figure}

The observed behavior for $h_{E}$ correlates with the changes in the contribution of the interfacial surface spins to the hysteresis loops displayed in Fig. \ref{Fig.7}(b), where it can be seen that the change in the fraction of pinned spins decreases for the particles with thinner shells (black, red and green curves) as compared to the one giving maximum EB (blue curve). Similar trend is observed for the $h_C$ curve, which can be understood by noticing that the coercive field, at difference of $h_{E}$, is directly related to the reversal of the core spins. As can be observed in Fig. \ref{Fig.10}, where a snapshot of the local changes in spin orientations between the positive and negative coercive field is depicted. The reversal of the core drags some of the inner spins at the interface, which explains the increase of $h_C$ with $D_{core}$. In contrast, spins at the outer part of the interface remain pinned, thus contributing to the loop shift.  Finally, let us also remark that the results of the hysteresis loops at finite temperatures  for sample with $t_{shell} = 17.5a $ shown in Figs. \ref{Fig.9}(c, d) indicate a thermal dependence  $h_{E}$  and $h_C$ which is in agreement with the monotonous decrease also observed experimentally. 

\section{Conclusions}
We report synthesis of Co based NP in the silica matrix. The controlled oxidation leads to the Co core with the Co$_3$O$_4$ shell structure. The core-shell size is varied keeping fixed overall size of the particle and negligible interparticle interaction is maintained  through the dispersion of Co/Co$_3$O$_4$ NP in the silica matrix of 10 \% volume fraction. Absence of trace amount of other oxide phase such as, CoO, is confirmed from XRD, TEM, and XPS analysis. Although the parent oxide has much lower ordering temperature, we have reported the existence of EB bias effects that persist up to almost room temperature. The maximum EB is observed for the samples with intermediate shell thickness that is accompanied by the maximum value of the  interfacial strain. The experimental results have been complimented with simulations based on atomistic spin model of individual NP with realistic sizes. The results of simulations reproduce qualitatively the observed EB phenomenology and suggest that interface pinning mechanism directs the EB effect and its dependence on the specific material parameters as well as geometry of the NP. 

\begin{acknowledgements}
S.G. wishes to thank DST, India (Project No. SB/S2/CMP-029/2014) for the financial support. SQUID magnetometer of Quantum Design and TEM are used in this study under DST the project, Unit of Nanoscience at Indian Association for the Cultivation of Science, Jadavpur, India. O. I. acknowledges financial support form the Spanish MINECO (MAT2012-33037, MAT2015-68772-P (MINECO/FEDER)), Catalan DURSI (2014SGR220) and European Union FEDER Funds (Una manera de hacer Europa), also CSUC for supercomputer facilities.
\end{acknowledgements}
%


\begin{thebibliography}{55}%
\makeatletter
\providecommand \@ifxundefined [1]{%
 \@ifx{#1\undefined}
}%
\providecommand \@ifnum [1]{%
 \ifnum #1\expandafter \@firstoftwo
 \else \expandafter \@secondoftwo
 \fi
}%
\providecommand \@ifx [1]{%
 \ifx #1\expandafter \@firstoftwo
 \else \expandafter \@secondoftwo
 \fi
}%
\providecommand \natexlab [1]{#1}%
\providecommand \enquote  [1]{``#1''}%
\providecommand \bibnamefont  [1]{#1}%
\providecommand \bibfnamefont [1]{#1}%
\providecommand \citenamefont [1]{#1}%
\providecommand \href@noop [0]{\@secondoftwo}%
\providecommand \href [0]{\begingroup \@sanitize@url \@href}%
\providecommand \@href[1]{\@@startlink{#1}\@@href}%
\providecommand \@@href[1]{\endgroup#1\@@endlink}%
\providecommand \@sanitize@url [0]{\catcode `\\12\catcode `\$12\catcode
  `\&12\catcode `\#12\catcode `\^12\catcode `\_12\catcode `\%12\relax}%
\providecommand \@@startlink[1]{}%
\providecommand \@@endlink[0]{}%
\providecommand \url  [0]{\begingroup\@sanitize@url \@url }%
\providecommand \@url [1]{\endgroup\@href {#1}{\urlprefix }}%
\providecommand \urlprefix  [0]{URL }%
\providecommand \Eprint [0]{\href }%
\providecommand \doibase [0]{http://dx.doi.org/}%
\providecommand \selectlanguage [0]{\@gobble}%
\providecommand \bibinfo  [0]{\@secondoftwo}%
\providecommand \bibfield  [0]{\@secondoftwo}%
\providecommand \translation [1]{[#1]}%
\providecommand \BibitemOpen [0]{}%
\providecommand \bibitemStop [0]{}%
\providecommand \bibitemNoStop [0]{.\EOS\space}%
\providecommand \EOS [0]{\spacefactor3000\relax}%
\providecommand \BibitemShut  [1]{\csname bibitem#1\endcsname}%
\let\auto@bib@innerbib\@empty
\bibitem [{\citenamefont {Dormann}\ \emph {et~al.}(1997)\citenamefont
  {Dormann}, \citenamefont {Fiorani},\ and\ \citenamefont
  {Tronc}}]{Dormann_ACP97}%
  \BibitemOpen
  \bibfield  {author} {\bibinfo {author} {\bibfnamefont {J.~L.}\ \bibnamefont
  {Dormann}}, \bibinfo {author} {\bibfnamefont {D.}~\bibnamefont {Fiorani}}, \
  and\ \bibinfo {author} {\bibfnamefont {E.}~\bibnamefont {Tronc}},\ }\href
  {\doibase 10.1002/9780470141571.ch4} {\bibfield  {journal} {\bibinfo
  {journal} {Adv. Chem. Phys.}\ }\textbf {\bibinfo {volume} {98}},\ \bibinfo
  {pages} {283} (\bibinfo {year} {1997})}\BibitemShut {NoStop}%
\bibitem [{\citenamefont {Kodama}(1999)}]{Kodama_jmmm1999}%
  \BibitemOpen
  \bibfield  {author} {\bibinfo {author} {\bibfnamefont {R.}~\bibnamefont
  {Kodama}},\ }\href {\doibase 10.1016/S0304-8853(99)00347-9} {\bibfield
  {journal} {\bibinfo  {journal} {J. Magn .Magn. Mater.}\ }\textbf {\bibinfo
  {volume} {200}},\ \bibinfo {pages} {359 } (\bibinfo {year}
  {1999})}\BibitemShut {NoStop}%
\bibitem [{\citenamefont {Fiorani}(2005)}]{Fiorani_book2005}%
  \BibitemOpen
  \bibfield  {author} {\bibinfo {author} {\bibfnamefont {D.}~\bibnamefont
  {Fiorani}},\ }\href {\doibase 10.1007/b136494} {\emph {\bibinfo {title}
  {Surface Effects in Magnetic Nanoparticles}}}\ (\bibinfo  {publisher}
  {Springer US},\ \bibinfo {address} {Boston, MA},\ \bibinfo {year}
  {2005})\BibitemShut {NoStop}%
\bibitem [{\citenamefont {Iglesias}\ and\ \citenamefont
  {Labarta}(2001)}]{Iglesias_prb2001}%
  \BibitemOpen
  \bibfield  {author} {\bibinfo {author} {\bibfnamefont {O.}~\bibnamefont
  {Iglesias}}\ and\ \bibinfo {author} {\bibfnamefont {A.}~\bibnamefont
  {Labarta}},\ }\href {\doibase 10.1103/PhysRevB.63.184416} {\bibfield
  {journal} {\bibinfo  {journal} {Phys. Rev. B}\ }\textbf {\bibinfo {volume}
  {63}},\ \bibinfo {pages} {184416} (\bibinfo {year} {2001})}\BibitemShut
  {NoStop}%
\bibitem [{\citenamefont {Sabsabi}\ \emph {et~al.}(2013)\citenamefont
  {Sabsabi}, \citenamefont {Vernay}, \citenamefont {Iglesias},\ and\
  \citenamefont {Kachkachi}}]{Sabsabi_prb2013}%
  \BibitemOpen
  \bibfield  {author} {\bibinfo {author} {\bibfnamefont {Z.}~\bibnamefont
  {Sabsabi}}, \bibinfo {author} {\bibfnamefont {F.}~\bibnamefont {Vernay}},
  \bibinfo {author} {\bibfnamefont {O.}~\bibnamefont {Iglesias}}, \ and\
  \bibinfo {author} {\bibfnamefont {H.}~\bibnamefont {Kachkachi}},\ }\href
  {\doibase 10.1103/PhysRevB.88.104424} {\bibfield  {journal} {\bibinfo
  {journal} {Phys. Rev. B}\ }\textbf {\bibinfo {volume} {88}},\ \bibinfo
  {pages} {104424} (\bibinfo {year} {2013})}\BibitemShut {NoStop}%
\bibitem [{\citenamefont {Nogu\'{e}s}\ and\ \citenamefont
  {Schuller}(1999)}]{Nogues_JMMM1999}%
  \BibitemOpen
  \bibfield  {author} {\bibinfo {author} {\bibfnamefont {J.}~\bibnamefont
  {Nogu\'{e}s}}\ and\ \bibinfo {author} {\bibfnamefont {I.~K.}\ \bibnamefont
  {Schuller}},\ }\href {\doibase 10.1016/S0304-8853(98)00266-2} {\bibfield
  {journal} {\bibinfo  {journal} {J. Magn .Magn. Mater.}\ }\textbf {\bibinfo
  {volume} {192}},\ \bibinfo {pages} {203 } (\bibinfo {year}
  {1999})}\BibitemShut {NoStop}%
\bibitem [{\citenamefont {Nogués}\ \emph {et~al.}(2005)\citenamefont
  {Nogués}, \citenamefont {Sort}, \citenamefont {Langlais}, \citenamefont
  {Skumryev}, \citenamefont {Suriñach}, \citenamefont {Muñoz},\ and\
  \citenamefont {Baró}}]{Nogues_physrep2005}%
  \BibitemOpen
  \bibfield  {author} {\bibinfo {author} {\bibfnamefont {J.}~\bibnamefont
  {Nogués}}, \bibinfo {author} {\bibfnamefont {J.}~\bibnamefont {Sort}},
  \bibinfo {author} {\bibfnamefont {V.}~\bibnamefont {Langlais}}, \bibinfo
  {author} {\bibfnamefont {V.}~\bibnamefont {Skumryev}}, \bibinfo {author}
  {\bibfnamefont {S.}~\bibnamefont {Suriñach}}, \bibinfo {author}
  {\bibfnamefont {J.}~\bibnamefont {Muñoz}}, \ and\ \bibinfo {author}
  {\bibfnamefont {M.}~\bibnamefont {Baró}},\ }\href {\doibase
  dx.doi.org/10.1016/j.physrep.2005.08.004} {\bibfield  {journal} {\bibinfo
  {journal} {Physics Reports}\ }\textbf {\bibinfo {volume} {422}},\ \bibinfo
  {pages} {65 } (\bibinfo {year} {2005})}\BibitemShut {NoStop}%
\bibitem [{\citenamefont {Iglesias}\ \emph {et~al.}(2008)\citenamefont
  {Iglesias}, \citenamefont {Labarta},\ and\ \citenamefont
  {Batlle}}]{Iglesias_jnn2008}%
  \BibitemOpen
  \bibfield  {author} {\bibinfo {author} {\bibfnamefont {O.}~\bibnamefont
  {Iglesias}}, \bibinfo {author} {\bibfnamefont {A.}~\bibnamefont {Labarta}}, \
  and\ \bibinfo {author} {\bibfnamefont {X.}~\bibnamefont {Batlle}},\ }\href
  {\doibase 10.1166/jnn.2008.015} {\bibfield  {journal} {\bibinfo  {journal}
  {J. Nanosc. Nanotech.}\ }\textbf {\bibinfo {volume} {8}},\ \bibinfo {pages}
  {2761} (\bibinfo {year} {2008})}\BibitemShut {NoStop}%
\bibitem [{\citenamefont {Giri}\ \emph {et~al.}(2011)\citenamefont {Giri},
  \citenamefont {Patra},\ and\ \citenamefont {Majumdar}}]{Giri_jpcm2011}%
  \BibitemOpen
  \bibfield  {author} {\bibinfo {author} {\bibfnamefont {S.}~\bibnamefont
  {Giri}}, \bibinfo {author} {\bibfnamefont {M.}~\bibnamefont {Patra}}, \ and\
  \bibinfo {author} {\bibfnamefont {S.}~\bibnamefont {Majumdar}},\ }\href
  {\doibase 10.1088/0953-8984/23/7/073201} {\bibfield  {journal} {\bibinfo
  {journal} {J. Phys.: Condens. Matter}\ }\textbf {\bibinfo {volume} {23}},\
  \bibinfo {pages} {073201} (\bibinfo {year} {2011})}\BibitemShut {NoStop}%
\bibitem [{\citenamefont {Manna}\ and\ \citenamefont
  {Yusuf}(2014)}]{Manna_PhysRep2014}%
  \BibitemOpen
  \bibfield  {author} {\bibinfo {author} {\bibfnamefont {P.}~\bibnamefont
  {Manna}}\ and\ \bibinfo {author} {\bibfnamefont {S.}~\bibnamefont {Yusuf}},\
  }\href {\doibase 10.1016/j.physrep.2013.10.002} {\bibfield  {journal}
  {\bibinfo  {journal} {Physics Reports}\ }\textbf {\bibinfo {volume} {535}},\
  \bibinfo {pages} {61 } (\bibinfo {year} {2014})}\BibitemShut {NoStop}%
\bibitem [{\citenamefont {Gierlings}\ \emph {et~al.}(2002)\citenamefont
  {Gierlings}, \citenamefont {Prandolini}, \citenamefont {Fritzsche},
  \citenamefont {Gruyters},\ and\ \citenamefont {Riegel}}]{Gierlings_PRB2002}%
  \BibitemOpen
  \bibfield  {author} {\bibinfo {author} {\bibfnamefont {M.}~\bibnamefont
  {Gierlings}}, \bibinfo {author} {\bibfnamefont {M.~J.}\ \bibnamefont
  {Prandolini}}, \bibinfo {author} {\bibfnamefont {H.}~\bibnamefont
  {Fritzsche}}, \bibinfo {author} {\bibfnamefont {M.}~\bibnamefont {Gruyters}},
  \ and\ \bibinfo {author} {\bibfnamefont {D.}~\bibnamefont {Riegel}},\ }\href
  {\doibase 10.1103/PhysRevB.65.092407} {\bibfield  {journal} {\bibinfo
  {journal} {Phys. Rev. B}\ }\textbf {\bibinfo {volume} {65}},\ \bibinfo
  {pages} {092407} (\bibinfo {year} {2002})}\BibitemShut {NoStop}%
\bibitem [{\citenamefont {Skumryev}\ \emph {et~al.}(2003)\citenamefont
  {Skumryev}, \citenamefont {Stoyanov}, \citenamefont {Zhang}, \citenamefont
  {Hadjipanayis}, \citenamefont {Givord},\ and\ \citenamefont
  {Nogu{\'{e}}s}}]{Skumryev_Nature2003}%
  \BibitemOpen
  \bibfield  {author} {\bibinfo {author} {\bibfnamefont {V.}~\bibnamefont
  {Skumryev}}, \bibinfo {author} {\bibfnamefont {S.}~\bibnamefont {Stoyanov}},
  \bibinfo {author} {\bibfnamefont {Y.}~\bibnamefont {Zhang}}, \bibinfo
  {author} {\bibfnamefont {G.}~\bibnamefont {Hadjipanayis}}, \bibinfo {author}
  {\bibfnamefont {D.}~\bibnamefont {Givord}}, \ and\ \bibinfo {author}
  {\bibfnamefont {J.}~\bibnamefont {Nogu{\'{e}}s}},\ }\href {\doibase
  10.1038/nature01687} {\bibfield  {journal} {\bibinfo  {journal} {Nature}\
  }\textbf {\bibinfo {volume} {422}},\ \bibinfo {pages} {850} (\bibinfo {year}
  {2003})}\BibitemShut {NoStop}%
\bibitem [{\citenamefont {te~Velthuis}\ \emph {et~al.}(2000)\citenamefont
  {te~Velthuis}, \citenamefont {Berger}, \citenamefont {Felcher}, \citenamefont
  {Hill},\ and\ \citenamefont {Dahlberg}}]{Velthuis_jap2000}%
  \BibitemOpen
  \bibfield  {author} {\bibinfo {author} {\bibfnamefont {S.~G.~E.}\
  \bibnamefont {te~Velthuis}}, \bibinfo {author} {\bibfnamefont
  {A.}~\bibnamefont {Berger}}, \bibinfo {author} {\bibfnamefont {G.~P.}\
  \bibnamefont {Felcher}}, \bibinfo {author} {\bibfnamefont {B.~K.}\
  \bibnamefont {Hill}}, \ and\ \bibinfo {author} {\bibfnamefont {E.~D.}\
  \bibnamefont {Dahlberg}},\ }\href {\doibase dx.doi.org/10.1063/1.373243}
  {\bibfield  {journal} {\bibinfo  {journal} {J. Appl. Phys.}\ }\textbf
  {\bibinfo {volume} {87}},\ \bibinfo {pages} {5046} (\bibinfo {year}
  {2000})}\BibitemShut {NoStop}%
\bibitem [{\citenamefont {Radu}\ \emph {et~al.}(2003)\citenamefont {Radu},
  \citenamefont {Etzkorn}, \citenamefont {Siebrecht}, \citenamefont {Schmitte},
  \citenamefont {Westerholt},\ and\ \citenamefont {Zabel}}]{Radu_prb2003}%
  \BibitemOpen
  \bibfield  {author} {\bibinfo {author} {\bibfnamefont {F.}~\bibnamefont
  {Radu}}, \bibinfo {author} {\bibfnamefont {M.}~\bibnamefont {Etzkorn}},
  \bibinfo {author} {\bibfnamefont {R.}~\bibnamefont {Siebrecht}}, \bibinfo
  {author} {\bibfnamefont {T.}~\bibnamefont {Schmitte}}, \bibinfo {author}
  {\bibfnamefont {K.}~\bibnamefont {Westerholt}}, \ and\ \bibinfo {author}
  {\bibfnamefont {H.}~\bibnamefont {Zabel}},\ }\href {\doibase
  10.1103/PhysRevB.67.134409} {\bibfield  {journal} {\bibinfo  {journal} {Phys.
  Rev. B}\ }\textbf {\bibinfo {volume} {67}},\ \bibinfo {pages} {134409}
  (\bibinfo {year} {2003})}\BibitemShut {NoStop}%
\bibitem [{\citenamefont {Das}\ \emph {et~al.}(2009)\citenamefont {Das},
  \citenamefont {Patra}, \citenamefont {Majumdar},\ and\ \citenamefont
  {Giri}}]{Das_jac2009}%
  \BibitemOpen
  \bibfield  {author} {\bibinfo {author} {\bibfnamefont {S.}~\bibnamefont
  {Das}}, \bibinfo {author} {\bibfnamefont {M.}~\bibnamefont {Patra}}, \bibinfo
  {author} {\bibfnamefont {S.}~\bibnamefont {Majumdar}}, \ and\ \bibinfo
  {author} {\bibfnamefont {S.}~\bibnamefont {Giri}},\ }\href {\doibase
  10.1016/j.jallcom.2009.08.143} {\bibfield  {journal} {\bibinfo  {journal}
  {Journal of Alloys and Compounds}\ }\textbf {\bibinfo {volume} {488}},\
  \bibinfo {pages} {27} (\bibinfo {year} {2009})}\BibitemShut {NoStop}%
\bibitem [{\citenamefont {Wu}\ \emph {et~al.}(2007)\citenamefont {Wu},
  \citenamefont {Li},\ and\ \citenamefont {Liu}}]{Wu_jpcm2007}%
  \BibitemOpen
  \bibfield  {author} {\bibinfo {author} {\bibfnamefont {M.~H.}\ \bibnamefont
  {Wu}}, \bibinfo {author} {\bibfnamefont {Q.~C.}\ \bibnamefont {Li}}, \ and\
  \bibinfo {author} {\bibfnamefont {J.-M.}\ \bibnamefont {Liu}},\ }\href
  {http://stacks.iop.org/0953-8984/19/i=18/a=186202} {\bibfield  {journal}
  {\bibinfo  {journal} {J. Phys.: Condens. Matter}\ }\textbf {\bibinfo {volume}
  {19}},\ \bibinfo {pages} {186202} (\bibinfo {year} {2007})}\BibitemShut
  {NoStop}%
\bibitem [{\citenamefont {Chappert}\ \emph {et~al.}(2007)\citenamefont
  {Chappert}, \citenamefont {Fert},\ and\ \citenamefont {{Van
  Dau}}}]{Chappert_NatMater2007}%
  \BibitemOpen
  \bibfield  {author} {\bibinfo {author} {\bibfnamefont {C.}~\bibnamefont
  {Chappert}}, \bibinfo {author} {\bibfnamefont {A.}~\bibnamefont {Fert}}, \
  and\ \bibinfo {author} {\bibfnamefont {F.~N.}\ \bibnamefont {{Van Dau}}},\
  }\href {\doibase 10.1038/nmat2024} {\bibfield  {journal} {\bibinfo  {journal}
  {Nature Materials}\ }\textbf {\bibinfo {volume} {6}},\ \bibinfo {pages} {813}
  (\bibinfo {year} {2007})}\BibitemShut {NoStop}%
\bibitem [{\citenamefont {L{\'{o}}pez-Ortega}\ \emph
  {et~al.}(2015)\citenamefont {L{\'{o}}pez-Ortega}, \citenamefont {Estrader},
  \citenamefont {Salazar-Alvarez}, \citenamefont {Roca},\ and\ \citenamefont
  {Nogu{\'{e}}s}}]{Lopez-Ortega_PhysRep2015}%
  \BibitemOpen
  \bibfield  {author} {\bibinfo {author} {\bibfnamefont {A.}~\bibnamefont
  {L{\'{o}}pez-Ortega}}, \bibinfo {author} {\bibfnamefont {M.}~\bibnamefont
  {Estrader}}, \bibinfo {author} {\bibfnamefont {G.}~\bibnamefont
  {Salazar-Alvarez}}, \bibinfo {author} {\bibfnamefont {A.~G.}\ \bibnamefont
  {Roca}}, \ and\ \bibinfo {author} {\bibfnamefont {J.}~\bibnamefont
  {Nogu{\'{e}}s}},\ }\href {\doibase 10.1016/j.physrep.2014.09.007} {\bibfield
  {journal} {\bibinfo  {journal} {Physics Reports}\ }\textbf {\bibinfo {volume}
  {553}},\ \bibinfo {pages} {1} (\bibinfo {year} {2015})}\BibitemShut {NoStop}%
\bibitem [{\citenamefont {O'Handley}(2000)}]{OHandley_2000}%
  \BibitemOpen
  \bibfield  {author} {\bibinfo {author} {\bibfnamefont {R.~C.}\ \bibnamefont
  {O'Handley}},\ }\href@noop {} {\emph {\bibinfo {title} {Modern Magnetic
  Materials: Principles and Applications}}}\ (\bibinfo  {publisher} {John Wiley
  and Sons},\ \bibinfo {address} {New York},\ \bibinfo {year}
  {2000})\BibitemShut {NoStop}%
\bibitem [{\citenamefont {Haneda}\ and\ \citenamefont
  {Morrish}(1979)}]{Haneda_Nature1979}%
  \BibitemOpen
  \bibfield  {author} {\bibinfo {author} {\bibfnamefont {K.}~\bibnamefont
  {Haneda}}\ and\ \bibinfo {author} {\bibfnamefont {A.~H.}\ \bibnamefont
  {Morrish}},\ }\href {\doibase 10.1038/282186a0} {\bibfield  {journal}
  {\bibinfo  {journal} {Nature}\ }\textbf {\bibinfo {volume} {282}},\ \bibinfo
  {pages} {186} (\bibinfo {year} {1979})}\BibitemShut {NoStop}%
\bibitem [{\citenamefont {Giri}\ \emph {et~al.}(2001)\citenamefont {Giri},
  \citenamefont {Ganguli},\ and\ \citenamefont {Bhattacharya}}]{Giri_ASC2001}%
  \BibitemOpen
  \bibfield  {author} {\bibinfo {author} {\bibfnamefont {S.}~\bibnamefont
  {Giri}}, \bibinfo {author} {\bibfnamefont {S.}~\bibnamefont {Ganguli}}, \
  and\ \bibinfo {author} {\bibfnamefont {M.}~\bibnamefont {Bhattacharya}},\
  }\href {\doibase http://dx.doi.org/10.1016/S0169-4332(01)00446-9} {\bibfield
  {journal} {\bibinfo  {journal} {Applied Surface Science}\ }\textbf {\bibinfo
  {volume} {182}},\ \bibinfo {pages} {345 } (\bibinfo {year} {2001})},\
  \bibinfo {note} {proceedings of the International Workshop on
  Nanomaterials}\BibitemShut {NoStop}%
\bibitem [{\citenamefont {Goodenough}(1963)}]{Goodenough_book1963}%
  \BibitemOpen
  \bibfield  {author} {\bibinfo {author} {\bibfnamefont {J.~B.}\ \bibnamefont
  {Goodenough}},\ }\href@noop {} {\emph {\bibinfo {title} {Magnetism and the
  Chemical Bond}}}\ (\bibinfo  {publisher} {Interscience Publishers},\ \bibinfo
  {address} {New York},\ \bibinfo {year} {1963})\BibitemShut {NoStop}%
\bibitem [{\citenamefont {Meiklejohn}\ and\ \citenamefont
  {Bean}(1956)}]{Meiklejohn_prb56}%
  \BibitemOpen
  \bibfield  {author} {\bibinfo {author} {\bibfnamefont {W.}~\bibnamefont
  {Meiklejohn}}\ and\ \bibinfo {author} {\bibfnamefont {C.}~\bibnamefont
  {Bean}},\ }\href {\doibase 10.1103/PhysRev.102.1413} {\bibfield  {journal}
  {\bibinfo  {journal} {Physical Review}\ }\textbf {\bibinfo {volume} {102}},\
  \bibinfo {pages} {1413} (\bibinfo {year} {1956})}\BibitemShut {NoStop}%
\bibitem [{\citenamefont {Fontai\~{n}a Troiti\~{n}o}\ \emph
  {et~al.}(2014)\citenamefont {Fontai\~{n}a Troiti\~{n}o}, \citenamefont
  {Liébana-Vi\~{n}as}, \citenamefont {Rodríguez-González}, \citenamefont
  {Kovács}, \citenamefont {Li}, , \citenamefont {Spasova}, \citenamefont
  {Farle},\ and\ \citenamefont {Salgueiri\~{n}o}}]{Fontaina_NL2004}%
  \BibitemOpen
  \bibfield  {author} {\bibinfo {author} {\bibfnamefont {N.}~\bibnamefont
  {Fontai\~{n}a Troiti\~{n}o}}, \bibinfo {author} {\bibfnamefont
  {S.}~\bibnamefont {Liébana-Vi\~{n}as}}, \bibinfo {author} {\bibfnamefont
  {B.}~\bibnamefont {Rodríguez-González}}, \bibinfo {author} {\bibfnamefont
  {A.}~\bibnamefont {Kovács}}, \bibinfo {author} {\bibfnamefont {Z.-A.}\
  \bibnamefont {Li}}, , \bibinfo {author} {\bibfnamefont {M.}~\bibnamefont
  {Spasova}}, \bibinfo {author} {\bibfnamefont {M.}~\bibnamefont {Farle}}, \
  and\ \bibinfo {author} {\bibfnamefont {V.}~\bibnamefont {Salgueiri\~{n}o}},\
  }\href {\doibase 10.1021/nl4038533} {\bibfield  {journal} {\bibinfo
  {journal} {Nano Letters}\ }\textbf {\bibinfo {volume} {14}},\ \bibinfo
  {pages} {640} (\bibinfo {year} {2014})}\BibitemShut {NoStop}%
\bibitem [{\citenamefont {Li}\ \emph {et~al.}(2015)\citenamefont {Li},
  \citenamefont {Fontai\~{n}a Troiti\~{n}o}, \citenamefont {Kovács},
  \citenamefont {Liébana-Vi\~{n}as}, \citenamefont {Spasova}, \citenamefont
  {Dunin-Borkowski}, \citenamefont {Müller}, \citenamefont {Doennig},
  \citenamefont {Pentcheva}, \citenamefont {Farle},\ and\ \citenamefont
  {Salgueiri\~{n}o}}]{Li_SciRep2015}%
  \BibitemOpen
  \bibfield  {author} {\bibinfo {author} {\bibfnamefont {Z.-A.}\ \bibnamefont
  {Li}}, \bibinfo {author} {\bibfnamefont {N.}~\bibnamefont {Fontai\~{n}a
  Troiti\~{n}o}}, \bibinfo {author} {\bibfnamefont {A.}~\bibnamefont
  {Kovács}}, \bibinfo {author} {\bibfnamefont {S.}~\bibnamefont
  {Liébana-Vi\~{n}as}}, \bibinfo {author} {\bibfnamefont {M.}~\bibnamefont
  {Spasova}}, \bibinfo {author} {\bibfnamefont {R.~E.}\ \bibnamefont
  {Dunin-Borkowski}}, \bibinfo {author} {\bibfnamefont {M.}~\bibnamefont
  {Müller}}, \bibinfo {author} {\bibfnamefont {D.}~\bibnamefont {Doennig}},
  \bibinfo {author} {\bibfnamefont {R.}~\bibnamefont {Pentcheva}}, \bibinfo
  {author} {\bibfnamefont {M.}~\bibnamefont {Farle}}, \ and\ \bibinfo {author}
  {\bibfnamefont {V.}~\bibnamefont {Salgueiri\~{n}o}},\ }\href {\doibase
  10.1038/srep07997} {\bibfield  {journal} {\bibinfo  {journal} {Scientific
  Reports}\ }\textbf {\bibinfo {volume} {5}},\ \bibinfo {pages} {7997}
  (\bibinfo {year} {2015})}\BibitemShut {NoStop}%
\bibitem [{\citenamefont {Simeonidis}\ \emph {et~al.}(2011)\citenamefont
  {Simeonidis}, \citenamefont {Martinez-Boubeta}, \citenamefont {Iglesias},
  \citenamefont {Cabot}, \citenamefont {Angelakeris}, \citenamefont
  {Mourdikoudis}, \citenamefont {Tsiaoussis}, \citenamefont {Delimitis},
  \citenamefont {Dendrinou-Samara},\ and\ \citenamefont
  {Kalogirou}}]{Simeonidis_prb2011}%
  \BibitemOpen
  \bibfield  {author} {\bibinfo {author} {\bibfnamefont {K.}~\bibnamefont
  {Simeonidis}}, \bibinfo {author} {\bibfnamefont {C.}~\bibnamefont
  {Martinez-Boubeta}}, \bibinfo {author} {\bibfnamefont {.}~\bibnamefont
  {Iglesias}}, \bibinfo {author} {\bibfnamefont {A.}~\bibnamefont {Cabot}},
  \bibinfo {author} {\bibfnamefont {M.}~\bibnamefont {Angelakeris}}, \bibinfo
  {author} {\bibfnamefont {S.}~\bibnamefont {Mourdikoudis}}, \bibinfo {author}
  {\bibfnamefont {I.}~\bibnamefont {Tsiaoussis}}, \bibinfo {author}
  {\bibfnamefont {A.}~\bibnamefont {Delimitis}}, \bibinfo {author}
  {\bibfnamefont {C.}~\bibnamefont {Dendrinou-Samara}}, \ and\ \bibinfo
  {author} {\bibfnamefont {O.}~\bibnamefont {Kalogirou}},\ }\href {\doibase
  10.1103/PhysRevB.84.144430} {\bibfield  {journal} {\bibinfo  {journal} {Phys.
  Rev. B}\ }\textbf {\bibinfo {volume} {84}},\ \bibinfo {pages} {144430}
  (\bibinfo {year} {2011})}\BibitemShut {NoStop}%
\bibitem [{\citenamefont {Wang}\ \emph
  {et~al.}(2005{\natexlab{a}})\citenamefont {Wang}, \citenamefont {You},
  \citenamefont {Tian}, \citenamefont {Wang}, \citenamefont {Sun},
  \citenamefont {Lu},\ and\ \citenamefont {Li}}]{Wang_ijmpb2005}%
  \BibitemOpen
  \bibfield  {author} {\bibinfo {author} {\bibfnamefont {Y.-X.}\ \bibnamefont
  {Wang}}, \bibinfo {author} {\bibfnamefont {B.}~\bibnamefont {You}}, \bibinfo
  {author} {\bibfnamefont {W.}~\bibnamefont {Tian}}, \bibinfo {author}
  {\bibfnamefont {Y.-X.}\ \bibnamefont {Wang}}, \bibinfo {author}
  {\bibfnamefont {L.}~\bibnamefont {Sun}}, \bibinfo {author} {\bibfnamefont
  {M.}~\bibnamefont {Lu}}, \ and\ \bibinfo {author} {\bibfnamefont
  {Q.}~\bibnamefont {Li}},\ }\href {\doibase 10.1142/S0217979205031353}
  {\bibfield  {journal} {\bibinfo  {journal} {Int. J. Mod. Phys. B}\ }\textbf
  {\bibinfo {volume} {19}},\ \bibinfo {pages} {2580} (\bibinfo {year}
  {2005}{\natexlab{a}})}\BibitemShut {NoStop}%
\bibitem [{\citenamefont {Wang}\ \emph
  {et~al.}(2005{\natexlab{b}})\citenamefont {Wang}, \citenamefont {Tian},
  \citenamefont {Wang}, \citenamefont {Sun}, \citenamefont {Li}, \citenamefont
  {You}, \citenamefont {Hu}, \citenamefont {Zhai},\ and\ \citenamefont
  {Lu}}]{Wang_ssc2005}%
  \BibitemOpen
  \bibfield  {author} {\bibinfo {author} {\bibfnamefont {Y.}~\bibnamefont
  {Wang}}, \bibinfo {author} {\bibfnamefont {W.}~\bibnamefont {Tian}}, \bibinfo
  {author} {\bibfnamefont {Y.}~\bibnamefont {Wang}}, \bibinfo {author}
  {\bibfnamefont {L.}~\bibnamefont {Sun}}, \bibinfo {author} {\bibfnamefont
  {Q.}~\bibnamefont {Li}}, \bibinfo {author} {\bibfnamefont {B.}~\bibnamefont
  {You}}, \bibinfo {author} {\bibfnamefont {A.}~\bibnamefont {Hu}}, \bibinfo
  {author} {\bibfnamefont {H.}~\bibnamefont {Zhai}}, \ and\ \bibinfo {author}
  {\bibfnamefont {M.}~\bibnamefont {Lu}},\ }\href {\doibase
  http://dx.doi.org/10.1016/j.ssc.2005.06.004} {\bibfield  {journal} {\bibinfo
  {journal} {Solid State Communications}\ }\textbf {\bibinfo {volume} {135}},\
  \bibinfo {pages} {725 } (\bibinfo {year} {2005}{\natexlab{b}})}\BibitemShut
  {NoStop}%
\bibitem [{\citenamefont {You}\ \emph {et~al.}(2003)\citenamefont {You},
  \citenamefont {Wang}, \citenamefont {Zhao}, \citenamefont {Sun},
  \citenamefont {Sheng}, \citenamefont {Pan}, \citenamefont {Du}, \citenamefont
  {Hu},\ and\ \citenamefont {Lu}}]{You_jap2003}%
  \BibitemOpen
  \bibfield  {author} {\bibinfo {author} {\bibfnamefont {B.}~\bibnamefont
  {You}}, \bibinfo {author} {\bibfnamefont {Y.}~\bibnamefont {Wang}}, \bibinfo
  {author} {\bibfnamefont {Y.}~\bibnamefont {Zhao}}, \bibinfo {author}
  {\bibfnamefont {L.}~\bibnamefont {Sun}}, \bibinfo {author} {\bibfnamefont
  {W.}~\bibnamefont {Sheng}}, \bibinfo {author} {\bibfnamefont
  {M.}~\bibnamefont {Pan}}, \bibinfo {author} {\bibfnamefont {J.}~\bibnamefont
  {Du}}, \bibinfo {author} {\bibfnamefont {A.}~\bibnamefont {Hu}}, \ and\
  \bibinfo {author} {\bibfnamefont {M.}~\bibnamefont {Lu}},\ }\href {\doibase
  10.1063/1.1543878} {\bibfield  {journal} {\bibinfo  {journal} {J. Appl.
  Phys.}\ }\textbf {\bibinfo {volume} {93}},\ \bibinfo {pages} {6587} (\bibinfo
  {year} {2003})}\BibitemShut {NoStop}%
\bibitem [{\citenamefont {Wang}\ \emph {et~al.}(2008)\citenamefont {Wang},
  \citenamefont {Zhang}, \citenamefont {Cao}, \citenamefont {Lu},\ and\
  \citenamefont {Yang}}]{Wang_jac2008}%
  \BibitemOpen
  \bibfield  {author} {\bibinfo {author} {\bibfnamefont {Y.}~\bibnamefont
  {Wang}}, \bibinfo {author} {\bibfnamefont {Y.}~\bibnamefont {Zhang}},
  \bibinfo {author} {\bibfnamefont {Y.}~\bibnamefont {Cao}}, \bibinfo {author}
  {\bibfnamefont {M.}~\bibnamefont {Lu}}, \ and\ \bibinfo {author}
  {\bibfnamefont {J.}~\bibnamefont {Yang}},\ }\href {\doibase
  http://dx.doi.org/10.1016/j.jallcom.2007.05.030} {\bibfield  {journal}
  {\bibinfo  {journal} {Journal of Alloys and Compounds}\ }\textbf {\bibinfo
  {volume} {450}},\ \bibinfo {pages} {128 } (\bibinfo {year}
  {2008})}\BibitemShut {NoStop}%
\bibitem [{\citenamefont {Ahmadvand}\ \emph {et~al.}(2015)\citenamefont
  {Ahmadvand}, \citenamefont {Safdari}, \citenamefont {Golikand}, \citenamefont
  {Dasgupta}, \citenamefont {Poddar},\ and\ \citenamefont
  {Salamati}}]{Ahmad_jmmm2015}%
  \BibitemOpen
  \bibfield  {author} {\bibinfo {author} {\bibfnamefont {H.}~\bibnamefont
  {Ahmadvand}}, \bibinfo {author} {\bibfnamefont {S.~R.}\ \bibnamefont
  {Safdari}}, \bibinfo {author} {\bibfnamefont {A.~N.}\ \bibnamefont
  {Golikand}}, \bibinfo {author} {\bibfnamefont {P.}~\bibnamefont {Dasgupta}},
  \bibinfo {author} {\bibfnamefont {A.}~\bibnamefont {Poddar}}, \ and\ \bibinfo
  {author} {\bibfnamefont {H.}~\bibnamefont {Salamati}},\ }\href {\doibase
  http://dx.doi.org/10.1016/j.jmmm.2014.10.021} {\bibfield  {journal} {\bibinfo
   {journal} {J. Magn. Magn. Mater.}\ }\textbf {\bibinfo {volume} {377}},\
  \bibinfo {pages} {19 } (\bibinfo {year} {2015})}\BibitemShut {NoStop}%
\bibitem [{\citenamefont {Bisht}\ and\ \citenamefont
  {Rajeev}(2011)}]{Bisht_scc2011}%
  \BibitemOpen
  \bibfield  {author} {\bibinfo {author} {\bibfnamefont {V.}~\bibnamefont
  {Bisht}}\ and\ \bibinfo {author} {\bibfnamefont {K.}~\bibnamefont {Rajeev}},\
  }\href {\doibase http://dx.doi.org/10.1016/j.ssc.2011.05.039} {\bibfield
  {journal} {\bibinfo  {journal} {Solid State Communications}\ }\textbf
  {\bibinfo {volume} {151}},\ \bibinfo {pages} {1275 } (\bibinfo {year}
  {2011})}\BibitemShut {NoStop}%
\bibitem [{\citenamefont {Ichiyanagi}\ and\ \citenamefont
  {Yamada}(2005)}]{Ichiyanagi_poly2005}%
  \BibitemOpen
  \bibfield  {author} {\bibinfo {author} {\bibfnamefont {Y.}~\bibnamefont
  {Ichiyanagi}}\ and\ \bibinfo {author} {\bibfnamefont {S.}~\bibnamefont
  {Yamada}},\ }\href {\doibase http://dx.doi.org/10.1016/j.poly.2005.03.158}
  {\bibfield  {journal} {\bibinfo  {journal} {Polyhedron}\ }\textbf {\bibinfo
  {volume} {24}},\ \bibinfo {pages} {2813 } (\bibinfo {year} {2005})},\
  \bibinfo {note} {proceedings of the 9th International Conference on
  Molecule-based Magnets (ICMM 2004)}\BibitemShut {NoStop}%
\bibitem [{\citenamefont {Dutta}\ \emph {et~al.}(2008)\citenamefont {Dutta},
  \citenamefont {Seehra}, \citenamefont {Thota},\ and\ \citenamefont
  {Kumar}}]{Dutta_jpcm2008}%
  \BibitemOpen
  \bibfield  {author} {\bibinfo {author} {\bibfnamefont {P.}~\bibnamefont
  {Dutta}}, \bibinfo {author} {\bibfnamefont {M.~S.}\ \bibnamefont {Seehra}},
  \bibinfo {author} {\bibfnamefont {S.}~\bibnamefont {Thota}}, \ and\ \bibinfo
  {author} {\bibfnamefont {J.}~\bibnamefont {Kumar}},\ }\href
  {http://stacks.iop.org/0953-8984/20/i=1/a=015218} {\bibfield  {journal}
  {\bibinfo  {journal} {J. Phys.: Condens. Matter}\ }\textbf {\bibinfo {volume}
  {20}},\ \bibinfo {pages} {015218} (\bibinfo {year} {2008})}\BibitemShut
  {NoStop}%
\bibitem [{\citenamefont {Silva}\ \emph {et~al.}(2010)\citenamefont {Silva},
  \citenamefont {Mill\'an}, \citenamefont {Palacio}, \citenamefont {Martins},
  \citenamefont {Trindade}, \citenamefont {Puente-Orench},\ and\ \citenamefont
  {Campo}}]{Silva_prb2010}%
  \BibitemOpen
  \bibfield  {author} {\bibinfo {author} {\bibfnamefont {N.~J.~O.}\
  \bibnamefont {Silva}}, \bibinfo {author} {\bibfnamefont {A.}~\bibnamefont
  {Mill\'an}}, \bibinfo {author} {\bibfnamefont {F.}~\bibnamefont {Palacio}},
  \bibinfo {author} {\bibfnamefont {M.}~\bibnamefont {Martins}}, \bibinfo
  {author} {\bibfnamefont {T.}~\bibnamefont {Trindade}}, \bibinfo {author}
  {\bibfnamefont {I.}~\bibnamefont {Puente-Orench}}, \ and\ \bibinfo {author}
  {\bibfnamefont {J.}~\bibnamefont {Campo}},\ }\href {\doibase
  10.1103/PhysRevB.82.094433} {\bibfield  {journal} {\bibinfo  {journal} {Phys.
  Rev. B}\ }\textbf {\bibinfo {volume} {82}},\ \bibinfo {pages} {094433}
  (\bibinfo {year} {2010})}\BibitemShut {NoStop}%
\bibitem [{\citenamefont {Cullity}\ and\ \citenamefont
  {Stock}(2004)}]{Cullity_book2004}%
  \BibitemOpen
  \bibfield  {author} {\bibinfo {author} {\bibfnamefont {B.}~\bibnamefont
  {Cullity}}\ and\ \bibinfo {author} {\bibfnamefont {S.}~\bibnamefont
  {Stock}},\ }\href@noop {} {\emph {\bibinfo {title} {Elements of X-ray
  Diffraction}}}\ (\bibinfo  {publisher} {Pearson},\ \bibinfo {address} {UK},\
  \bibinfo {year} {2004})\BibitemShut {NoStop}%
\bibitem [{\citenamefont {Zhuo}\ \emph {et~al.}(2009)\citenamefont {Zhuo},
  \citenamefont {Ge}, \citenamefont {Cao},\ and\ \citenamefont
  {Tang}}]{Zhuo_jcgd2009}%
  \BibitemOpen
  \bibfield  {author} {\bibinfo {author} {\bibfnamefont {L.}~\bibnamefont
  {Zhuo}}, \bibinfo {author} {\bibfnamefont {J.}~\bibnamefont {Ge}}, \bibinfo
  {author} {\bibfnamefont {L.}~\bibnamefont {Cao}}, \ and\ \bibinfo {author}
  {\bibfnamefont {B.}~\bibnamefont {Tang}},\ }\href {\doibase
  10.1021/cg070482r} {\bibfield  {journal} {\bibinfo  {journal} {Crystal Growth
  \& Design}\ }\textbf {\bibinfo {volume} {9}},\ \bibinfo {pages} {1} (\bibinfo
  {year} {2009})}\BibitemShut {NoStop}%
\bibitem [{\citenamefont {Ishida}(1983)}]{Nishizawa_bapd1983}%
  \BibitemOpen
  \bibfield  {author} {\bibinfo {author} {\bibfnamefont {T.~N.~K.}\
  \bibnamefont {Ishida}},\ }\href@noop {} {\bibfield  {journal} {\bibinfo
  {journal} {Journal of Phase Equilibria}\ }\textbf {\bibinfo {volume} {4}}
  (\bibinfo {year} {1983})}\BibitemShut {NoStop}%
\bibitem [{\citenamefont {Chandra}\ \emph {et~al.}(2012)\citenamefont
  {Chandra}, \citenamefont {Khurshid}, \citenamefont {Phan},\ and\
  \citenamefont {Srikanth}}]{Chandra_apl2012}%
  \BibitemOpen
  \bibfield  {author} {\bibinfo {author} {\bibfnamefont {S.}~\bibnamefont
  {Chandra}}, \bibinfo {author} {\bibfnamefont {H.}~\bibnamefont {Khurshid}},
  \bibinfo {author} {\bibfnamefont {M.-H.}\ \bibnamefont {Phan}}, \ and\
  \bibinfo {author} {\bibfnamefont {H.}~\bibnamefont {Srikanth}},\ }\href
  {\doibase 10.1063/1.4769350} {\bibfield  {journal} {\bibinfo  {journal}
  {Appl. Phys. Lett.}\ }\textbf {\bibinfo {volume} {101}},\ \bibinfo {pages}
  {232405} (\bibinfo {year} {2012})}\BibitemShut {NoStop}%
\bibitem [{\citenamefont {Feygenson}\ \emph {et~al.}(2010)\citenamefont
  {Feygenson}, \citenamefont {Yiu}, \citenamefont {Kou}, \citenamefont {Kim},\
  and\ \citenamefont {Aronson}}]{Feygenson_prb2010}%
  \BibitemOpen
  \bibfield  {author} {\bibinfo {author} {\bibfnamefont {M.}~\bibnamefont
  {Feygenson}}, \bibinfo {author} {\bibfnamefont {Y.}~\bibnamefont {Yiu}},
  \bibinfo {author} {\bibfnamefont {A.}~\bibnamefont {Kou}}, \bibinfo {author}
  {\bibfnamefont {K.-S.}\ \bibnamefont {Kim}}, \ and\ \bibinfo {author}
  {\bibfnamefont {M.~C.}\ \bibnamefont {Aronson}},\ }\href {\doibase
  10.1103/PhysRevB.81.195445} {\bibfield  {journal} {\bibinfo  {journal} {Phys.
  Rev. B}\ }\textbf {\bibinfo {volume} {81}},\ \bibinfo {pages} {195445}
  (\bibinfo {year} {2010})}\BibitemShut {NoStop}%
\bibitem [{\citenamefont {Roth}(1964)}]{Roth_jpcs1964}%
  \BibitemOpen
  \bibfield  {author} {\bibinfo {author} {\bibfnamefont {W.~L.}\ \bibnamefont
  {Roth}},\ }\href {\doibase 10.1016/0022-3697(64)90156-8} {\bibfield
  {journal} {\bibinfo  {journal} {J. Phys. Chem. Solids}\ }\textbf {\bibinfo
  {volume} {25}},\ \bibinfo {pages} {1} (\bibinfo {year} {1964})}\BibitemShut
  {NoStop}%
\bibitem [{\citenamefont {von Dreifus}\ \emph {et~al.}(2015)\citenamefont {von
  Dreifus}, \citenamefont {Pereira},\ and\ \citenamefont
  {de~Oliveira}}]{Dreifus_mre2015}%
  \BibitemOpen
  \bibfield  {author} {\bibinfo {author} {\bibfnamefont {D.}~\bibnamefont {von
  Dreifus}}, \bibinfo {author} {\bibfnamefont {E.~C.}\ \bibnamefont {Pereira}},
  \ and\ \bibinfo {author} {\bibfnamefont {A.~J.~A.}\ \bibnamefont
  {de~Oliveira}},\ }\href {\doibase 10.1088/2053-1591/2/11/116102} {\bibfield
  {journal} {\bibinfo  {journal} {Materials Research Express}\ }\textbf
  {\bibinfo {volume} {2}},\ \bibinfo {pages} {116102} (\bibinfo {year}
  {2015})}\BibitemShut {NoStop}%
\bibitem [{\citenamefont {Tracy}\ \emph {et~al.}(2005)\citenamefont {Tracy},
  \citenamefont {Weiss}, \citenamefont {Dinega},\ and\ \citenamefont
  {Bawendi}}]{Tracy_prb2005}%
  \BibitemOpen
  \bibfield  {author} {\bibinfo {author} {\bibfnamefont {J.~B.}\ \bibnamefont
  {Tracy}}, \bibinfo {author} {\bibfnamefont {D.~N.}\ \bibnamefont {Weiss}},
  \bibinfo {author} {\bibfnamefont {D.~P.}\ \bibnamefont {Dinega}}, \ and\
  \bibinfo {author} {\bibfnamefont {M.~G.}\ \bibnamefont {Bawendi}},\ }\href
  {\doibase 10.1103/PhysRevB.72.064404} {\bibfield  {journal} {\bibinfo
  {journal} {Phys. Rev. B}\ }\textbf {\bibinfo {volume} {72}},\ \bibinfo
  {pages} {064404} (\bibinfo {year} {2005})}\BibitemShut {NoStop}%
\bibitem [{\citenamefont {Tracy}\ and\ \citenamefont
  {Bawendi}(2006)}]{Tracy_prb2006}%
  \BibitemOpen
  \bibfield  {author} {\bibinfo {author} {\bibfnamefont {J.~B.}\ \bibnamefont
  {Tracy}}\ and\ \bibinfo {author} {\bibfnamefont {M.~G.}\ \bibnamefont
  {Bawendi}},\ }\href {\doibase 10.1103/PhysRevB.74.184434} {\bibfield
  {journal} {\bibinfo  {journal} {Phys. Rev. B}\ }\textbf {\bibinfo {volume}
  {74}},\ \bibinfo {pages} {184434} (\bibinfo {year} {2006})}\BibitemShut
  {NoStop}%
\bibitem [{\citenamefont {Rinaldi-Montes}\ \emph {et~al.}(2016)\citenamefont
  {Rinaldi-Montes}, \citenamefont {Gorria}, \citenamefont
  {Mart{\'{i}}nez-Blanco}, \citenamefont {Amghouz}, \citenamefont {Fuertes},
  \citenamefont {{Fern{\'{a}}ndez Barqu{\'{i}}n}}, \citenamefont
  {{Rodr{\'{i}}guez Fern{\'{a}}ndez}}, \citenamefont {Olivi}, \citenamefont
  {Aquilanti},\ and\ \citenamefont {Blanco}}]{Rinaldi-Montes_jmcc2016}%
  \BibitemOpen
  \bibfield  {author} {\bibinfo {author} {\bibfnamefont {N.}~\bibnamefont
  {Rinaldi-Montes}}, \bibinfo {author} {\bibfnamefont {P.}~\bibnamefont
  {Gorria}}, \bibinfo {author} {\bibfnamefont {D.}~\bibnamefont
  {Mart{\'{i}}nez-Blanco}}, \bibinfo {author} {\bibfnamefont {Z.}~\bibnamefont
  {Amghouz}}, \bibinfo {author} {\bibfnamefont {A.~B.}\ \bibnamefont
  {Fuertes}}, \bibinfo {author} {\bibfnamefont {L.}~\bibnamefont
  {{Fern{\'{a}}ndez Barqu{\'{i}}n}}}, \bibinfo {author} {\bibfnamefont
  {J.}~\bibnamefont {{Rodr{\'{i}}guez Fern{\'{a}}ndez}}}, \bibinfo {author}
  {\bibfnamefont {L.}~\bibnamefont {Olivi}}, \bibinfo {author} {\bibfnamefont
  {G.}~\bibnamefont {Aquilanti}}, \ and\ \bibinfo {author} {\bibfnamefont
  {J.~A.}\ \bibnamefont {Blanco}},\ }\href {\doibase 10.1039/C6TC00540C}
  {\bibfield  {journal} {\bibinfo  {journal} {J. Mater. Chem. C}\ }\textbf
  {\bibinfo {volume} {4}},\ \bibinfo {pages} {2302} (\bibinfo {year}
  {2016})}\BibitemShut {NoStop}%
\bibitem [{\citenamefont {Kovylina}\ \emph {et~al.}(2009)\citenamefont
  {Kovylina}, \citenamefont {del Muro}, \citenamefont {Konstantinovic},
  \citenamefont {Varela}, \citenamefont {Òscar Iglesias}, \citenamefont
  {Labarta},\ and\ \citenamefont {Batlle}}]{Kovylina_nanot2009}%
  \BibitemOpen
  \bibfield  {author} {\bibinfo {author} {\bibfnamefont {M.}~\bibnamefont
  {Kovylina}}, \bibinfo {author} {\bibfnamefont {M.~G.}\ \bibnamefont {del
  Muro}}, \bibinfo {author} {\bibfnamefont {Z.}~\bibnamefont {Konstantinovic}},
  \bibinfo {author} {\bibfnamefont {M.}~\bibnamefont {Varela}}, \bibinfo
  {author} {\bibnamefont {Òscar Iglesias}}, \bibinfo {author} {\bibfnamefont
  {A.}~\bibnamefont {Labarta}}, \ and\ \bibinfo {author} {\bibfnamefont
  {X.}~\bibnamefont {Batlle}},\ }\href
  {http://stacks.iop.org/0957-4484/20/i=17/a=175702} {\bibfield  {journal}
  {\bibinfo  {journal} {Nanotechnology}\ }\textbf {\bibinfo {volume} {20}},\
  \bibinfo {pages} {175702} (\bibinfo {year} {2009})}\BibitemShut {NoStop}%
\bibitem [{\citenamefont {Gangopadhyay}\ \emph {et~al.}(1993)\citenamefont
  {Gangopadhyay}, \citenamefont {Hadjipanayis}, \citenamefont {Sorensen},\ and\
  \citenamefont {Klabunde}}]{Gangopadhyay_jap1993}%
  \BibitemOpen
  \bibfield  {author} {\bibinfo {author} {\bibfnamefont {S.}~\bibnamefont
  {Gangopadhyay}}, \bibinfo {author} {\bibfnamefont {G.~C.}\ \bibnamefont
  {Hadjipanayis}}, \bibinfo {author} {\bibfnamefont {C.~M.}\ \bibnamefont
  {Sorensen}}, \ and\ \bibinfo {author} {\bibfnamefont {K.~J.}\ \bibnamefont
  {Klabunde}},\ }\href {\doibase 10.1063/1.352398} {\bibfield  {journal}
  {\bibinfo  {journal} {J. Appl. Phys.}\ }\textbf {\bibinfo {volume} {73}},\
  \bibinfo {pages} {6964} (\bibinfo {year} {1993})}\BibitemShut {NoStop}%
\bibitem [{\citenamefont {Chandra}\ \emph {et~al.}(2014)\citenamefont
  {Chandra}, \citenamefont {Huls}, \citenamefont {Phan}, \citenamefont
  {Srinath}, \citenamefont {Garcia}, \citenamefont {Lee}, \citenamefont {Wang},
  \citenamefont {Sun}, \citenamefont {Òscar Iglesias},\ and\ \citenamefont
  {Srikanth}}]{Chandra_Nano2014}%
  \BibitemOpen
  \bibfield  {author} {\bibinfo {author} {\bibfnamefont {S.}~\bibnamefont
  {Chandra}}, \bibinfo {author} {\bibfnamefont {N.~A.~F.}\ \bibnamefont
  {Huls}}, \bibinfo {author} {\bibfnamefont {M.~H.}\ \bibnamefont {Phan}},
  \bibinfo {author} {\bibfnamefont {S.}~\bibnamefont {Srinath}}, \bibinfo
  {author} {\bibfnamefont {M.~A.}\ \bibnamefont {Garcia}}, \bibinfo {author}
  {\bibfnamefont {Y.}~\bibnamefont {Lee}}, \bibinfo {author} {\bibfnamefont
  {C.}~\bibnamefont {Wang}}, \bibinfo {author} {\bibfnamefont {S.}~\bibnamefont
  {Sun}}, \bibinfo {author} {\bibnamefont {Òscar Iglesias}}, \ and\ \bibinfo
  {author} {\bibfnamefont {H.}~\bibnamefont {Srikanth}},\ }\href {\doibase
  10.1088/0957-4484/25/5/055702} {\bibfield  {journal} {\bibinfo  {journal}
  {Nanotechnology}\ }\textbf {\bibinfo {volume} {25}},\ \bibinfo {pages}
  {055702} (\bibinfo {year} {2014})}\BibitemShut {NoStop}%
\bibitem [{\citenamefont {Feygenson}\ \emph {et~al.}(2015)\citenamefont
  {Feygenson}, \citenamefont {Bauer}, \citenamefont {Gai}, \citenamefont
  {Marques}, \citenamefont {Aronson}, \citenamefont {Teng}, \citenamefont {Su},
  \citenamefont {Stanic}, \citenamefont {Urban}, \citenamefont {Beyer},\ and\
  \citenamefont {Dai}}]{Feygenson_prb2015}%
  \BibitemOpen
  \bibfield  {author} {\bibinfo {author} {\bibfnamefont {M.}~\bibnamefont
  {Feygenson}}, \bibinfo {author} {\bibfnamefont {J.~C.}\ \bibnamefont
  {Bauer}}, \bibinfo {author} {\bibfnamefont {Z.}~\bibnamefont {Gai}}, \bibinfo
  {author} {\bibfnamefont {C.}~\bibnamefont {Marques}}, \bibinfo {author}
  {\bibfnamefont {M.~C.}\ \bibnamefont {Aronson}}, \bibinfo {author}
  {\bibfnamefont {X.}~\bibnamefont {Teng}}, \bibinfo {author} {\bibfnamefont
  {D.}~\bibnamefont {Su}}, \bibinfo {author} {\bibfnamefont {V.}~\bibnamefont
  {Stanic}}, \bibinfo {author} {\bibfnamefont {V.~S.}\ \bibnamefont {Urban}},
  \bibinfo {author} {\bibfnamefont {K.~A.}\ \bibnamefont {Beyer}}, \ and\
  \bibinfo {author} {\bibfnamefont {S.}~\bibnamefont {Dai}},\ }\href {\doibase
  10.1103/PhysRevB.92.054416} {\bibfield  {journal} {\bibinfo  {journal} {Phys.
  Rev. B}\ }\textbf {\bibinfo {volume} {92}},\ \bibinfo {pages} {054416}
  (\bibinfo {year} {2015})}\BibitemShut {NoStop}%
\bibitem [{\citenamefont {{Del Bianco}}\ \emph {et~al.}(2004)\citenamefont
  {{Del Bianco}}, \citenamefont {Fiorani}, \citenamefont {Testa}, \citenamefont
  {Bonetti},\ and\ \citenamefont {Signorini}}]{DelBianco_prb2004}%
  \BibitemOpen
  \bibfield  {author} {\bibinfo {author} {\bibfnamefont {L.}~\bibnamefont {{Del
  Bianco}}}, \bibinfo {author} {\bibfnamefont {D.}~\bibnamefont {Fiorani}},
  \bibinfo {author} {\bibfnamefont {A.}~\bibnamefont {Testa}}, \bibinfo
  {author} {\bibfnamefont {E.}~\bibnamefont {Bonetti}}, \ and\ \bibinfo
  {author} {\bibfnamefont {L.}~\bibnamefont {Signorini}},\ }\href {\doibase
  10.1103/PhysRevB.70.052401} {\bibfield  {journal} {\bibinfo  {journal}
  {Physical Review B}\ }\textbf {\bibinfo {volume} {70}},\ \bibinfo {pages}
  {052401} (\bibinfo {year} {2004})}\BibitemShut {NoStop}%
\bibitem [{\citenamefont {Fiorani}\ \emph {et~al.}(2007)\citenamefont
  {Fiorani}, \citenamefont {Bianco}, \citenamefont {Testa},\ and\ \citenamefont
  {Trohidou}}]{Fiorani_jpcm2007}%
  \BibitemOpen
  \bibfield  {author} {\bibinfo {author} {\bibfnamefont {D.}~\bibnamefont
  {Fiorani}}, \bibinfo {author} {\bibfnamefont {L.~D.}\ \bibnamefont {Bianco}},
  \bibinfo {author} {\bibfnamefont {A.~M.}\ \bibnamefont {Testa}}, \ and\
  \bibinfo {author} {\bibfnamefont {K.~N.}\ \bibnamefont {Trohidou}},\ }\href
  {http://stacks.iop.org/0953-8984/19/i=22/a=225007} {\bibfield  {journal}
  {\bibinfo  {journal} {J. Phys.: Condens. Matter}\ }\textbf {\bibinfo {volume}
  {19}},\ \bibinfo {pages} {225007} (\bibinfo {year} {2007})}\BibitemShut
  {NoStop}%
\bibitem [{\citenamefont {Vasilakaki}\ and\ \citenamefont
  {Trohidou}(2009)}]{Vasilakaki_prb2009}%
  \BibitemOpen
  \bibfield  {author} {\bibinfo {author} {\bibfnamefont {M.}~\bibnamefont
  {Vasilakaki}}\ and\ \bibinfo {author} {\bibfnamefont {K.~N.}\ \bibnamefont
  {Trohidou}},\ }\href {\doibase 10.1103/PhysRevB.79.144402} {\bibfield
  {journal} {\bibinfo  {journal} {Phys. Rev. B}\ }\textbf {\bibinfo {volume}
  {79}},\ \bibinfo {pages} {144402} (\bibinfo {year} {2009})}\BibitemShut
  {NoStop}%
\bibitem [{\citenamefont {Cabot}\ \emph {et~al.}(2009)\citenamefont {Cabot},
  \citenamefont {Alivisatos}, \citenamefont {Puntes}, \citenamefont {Balcells},
  \citenamefont {Iglesias},\ and\ \citenamefont {Labarta}}]{Cabot_prb2009}%
  \BibitemOpen
  \bibfield  {author} {\bibinfo {author} {\bibfnamefont {A.}~\bibnamefont
  {Cabot}}, \bibinfo {author} {\bibfnamefont {A.~P.}\ \bibnamefont
  {Alivisatos}}, \bibinfo {author} {\bibfnamefont {V.~F.}\ \bibnamefont
  {Puntes}}, \bibinfo {author} {\bibfnamefont {L.}~\bibnamefont {Balcells}},
  \bibinfo {author} {\bibfnamefont {O.}~\bibnamefont {Iglesias}}, \ and\
  \bibinfo {author} {\bibfnamefont {A.}~\bibnamefont {Labarta}},\ }\href
  {\doibase 10.1103/PhysRevB.79.094419} {\bibfield  {journal} {\bibinfo
  {journal} {Phys. Rev. B}\ }\textbf {\bibinfo {volume} {79}},\ \bibinfo
  {pages} {094419} (\bibinfo {year} {2009})}\BibitemShut {NoStop}%
\bibitem [{\citenamefont {Balcells}\ \emph {et~al.}(2009)\citenamefont
  {Balcells}, \citenamefont {Mart\'{\i}nez}, \citenamefont {Iglesias},
  \citenamefont {Garc\'{\i}a-Mart\'{\i}n}, \citenamefont {Cebollada},
  \citenamefont {Garc\'{\i}a-Mart\'{\i}n}, \citenamefont {Armelles},
  \citenamefont {Sep\'ulveda},\ and\ \citenamefont
  {Alaverdyan}}]{Balcells_apl09}%
  \BibitemOpen
  \bibfield  {author} {\bibinfo {author} {\bibfnamefont {L.}~\bibnamefont
  {Balcells}}, \bibinfo {author} {\bibfnamefont {B.}~\bibnamefont
  {Mart\'{\i}nez}}, \bibinfo {author} {\bibfnamefont {{\O}.}~\bibnamefont
  {Iglesias}}, \bibinfo {author} {\bibfnamefont {J.~M.}\ \bibnamefont
  {Garc\'{\i}a-Mart\'{\i}n}}, \bibinfo {author} {\bibfnamefont
  {A.}~\bibnamefont {Cebollada}}, \bibinfo {author} {\bibfnamefont
  {A.}~\bibnamefont {Garc\'{\i}a-Mart\'{\i}n}}, \bibinfo {author}
  {\bibfnamefont {G.}~\bibnamefont {Armelles}}, \bibinfo {author}
  {\bibfnamefont {B.}~\bibnamefont {Sep\'ulveda}}, \ and\ \bibinfo {author}
  {\bibfnamefont {Y.}~\bibnamefont {Alaverdyan}},\ }\href {\doibase
  dx.doi.org/10.1063/1.3078411} {\bibfield  {journal} {\bibinfo  {journal}
  {Appl. Phys. Lett.}\ }\textbf {\bibinfo {volume} {94}},\ \bibinfo {pages}
  {062502} (\bibinfo {year} {2009})}\BibitemShut {NoStop}%
\bibitem [{\citenamefont {Iglesias}\ \emph {et~al.}(2005)\citenamefont
  {Iglesias}, \citenamefont {Batlle},\ and\ \citenamefont
  {Labarta}}]{Iglesias_prb2005}%
  \BibitemOpen
  \bibfield  {author} {\bibinfo {author} {\bibfnamefont {O.}~\bibnamefont
  {Iglesias}}, \bibinfo {author} {\bibfnamefont {X.}~\bibnamefont {Batlle}}, \
  and\ \bibinfo {author} {\bibfnamefont {A.}~\bibnamefont {Labarta}},\ }\href
  {\doibase 10.1103/PhysRevB.72.212401} {\bibfield  {journal} {\bibinfo
  {journal} {Phys. Rev. B}\ }\textbf {\bibinfo {volume} {72}},\ \bibinfo
  {pages} {212401} (\bibinfo {year} {2005})}\BibitemShut {NoStop}%
\end{thebibliography}
%

\end{document}